\documentclass[12pt,eqsecnum]{article}
\usepackage[dvips]{graphicx}
\usepackage{amssymb}
\usepackage{amsmath}
\usepackage{epstopdf}
\makeatletter

\@addtoreset{equation}{section}
\makeatletter

\newcommand{\psip}{{\psi^\prime}}
\newcommand{\psid}{\dot{\psi}}

\newcommand{\Pis}{P_\psi}

\newcommand{\g}{g_{(2)}}
\newcommand{\ma}[1]{\mbox{$\mathcal{#1}$}}

\newcommand{\D}{{\rm d}}

\newcommand{\dalm}{\kern1pt\vbox{\hrule height 0.9pt\hbox{\vrule width
0.9pt\hskip 2.5pt\vbox{\vskip 5.5pt}\hskip 3pt\vrule width 0.3pt}\hrule height
0.3pt}\kern1pt}

\def\b2hat{ {\hat b}_2 }

\def\none{\nonumber\\}

\def\be {\begin{equation}}
\def\ee  {\end{equation}}
\def\bea {\begin{eqnarray}}
\def\eea {\end{eqnarray}}

\begin{document}

\begin{titlepage}
\vfill
\begin{flushright}
\today
\end{flushright}

\vfill
\begin{center}
\baselineskip=16pt
{\Large\bf 
Hamiltonian dynamics of Lovelock black holes\\ with spherical symmetry\\
}
\vskip 0.5cm
{\large {\sl }}
\vskip 10.mm
{\bf Gabor Kunstatter${}^{a}$, Hideki Maeda${}^{b}$, and Tim Taves${}^{c}$} \\

\vskip 1cm
{
	${}^a$ Department of Physics, University of Winnipeg and Winnipeg Institute for Theoretical Physics, Winnipeg, Manitoba, Canada R3B 2E9\\
	${}^b$ Centro de Estudios Cient\'{\i}ficos (CECs), Casilla 1469, Valdivia, Chile. \\
	${}^c$ Department of Physics and Astronomy, University of Manitoba and Winnipeg Institute for Theoretical Physics, Winnipeg, Manitoba, Canada R3T 2N2\\
	\texttt{g.kunstatter-at-uwinnipeg.ca, hideki-at-cecs.cl, timtaves-at-gmail.com}

  }
\vspace{6pt}
\today
\end{center}
\vskip 0.2in
\par
\begin{center}
{\bf Abstract}
 \end{center}
\begin{quote}
We consider spherically symmetric black holes in generic Lovelock gravity. Using geometrodynamical variables we do a complete Hamiltonian analysis, including derivation of the super-Hamiltonian and super-momentum constraints and verification of suitable boundary conditions for asymptotically flat black holes. Our analysis leads to a remarkably simple fully reduced Hamiltonian for the vacuum gravitational sector that provides the starting point for the quantization of Lovelock block holes. Finally, we derive the completely reduced equations of motion for the collapse of a spherically symmetric, charged self-gravitating complex scalar field in generalized flat slice (Painlev\'{e}-Gullstrand) coordinates.
  \vfill
\vskip 2.mm
\end{quote}
\end{titlepage}




\tableofcontents

\newpage

\section{Introduction}
\label{Introduction}
The origin of the four-dimensionality of the present universe is one of the most fundamental problems in gravitational physics.
One possible explanation is that only the four-dimensional universe is stable in some physical sense and therefore chosen at the moment of its creation.
Another possibility is that after the creation of a higher-dimensional universe, the extra spatial dimensions other than our perceived three-dimensional space are compactified by some mechanism.
Einstein's general theory of relativity allows us to study this fundamental problem by setting the number, $n$, of spacetime dimensions as a tunable parameter.
It is therefore possible to look for critical values of $n$ beyond which spacetime properties change drastically. Such analyses can give us valuable insights about the origin of our four dimensional universe.

In this context, it is reasonable to focus on black holes because they are fundamental objects that encode many key features of the gravitational interaction. In four dimensions, asymptotically flat stationary black holes are characterized by a small number of parameters such as mass or angular momentum, somewhat analogous to atomic properties in chemistry.
This is a consequence of the black-hole uniqueness theorem asserting that the Kerr-Newman black hole is the unique asymptotically flat stationary and rotating black hole with a connected horizon in the Einstein-Maxwell system. (See~\cite{BHuniqueness} for review.) It is important to note that this uniqueness theorem is not valid in higher dimensions~\cite{er2008}, as explicitly shown in five dimensions, for example, by the existence of two distinct asymptotically flat black objects with the same mass and angular momentum but with different horizon topology~\cite{myersperry1986,er2002}. 
This fact already demonstrates one special feature of four-dimensional spacetime. $n=4$ is further singled out as a critical value in the framework of general relativity because asymptotically flat vacuum black holes do not exist in $n<4$ dimensions~\cite{ida2000}. 

When studying higher dimensions, it must be remembered that general relativity is not the only natural extension of four-dimensional Einstein gravity. General relativity is a quasi-linear second-order theory, which ensures the well-definedness of the initial value problem and the absence of ghosts. In 1971 Lovelock showed that in higher dimensions general relativity is just a special case of the most general class of theories satisfying these property. These more general theories are collectively called Lovelock gravity~\cite{Lovelock}.
The Lovelock Lagrangian consists of a sum of the dimensionally extended Euler densities in which the cosmological constant and the Einstein-Hilbert terms appear as the zeroth- and the first-order terms, respectively.
Just as the Einstein tensor trivially vanishes in two-dimensional spacetime, the second-order Lovelock Lagrangian (called the Gauss-Bonnet term) becomes purely topological in four spacetime dimensions and does not contribute to the field equations~\cite{lanczos}.
As a consequence, Lovelock gravity reduces to general relativity with a cosmological constant in four dimensions.

Motivation for studying Lovelock gravity is also provided by the fact that string/M-theory~\cite{string} requires the existence of extra spatial dimensions whereas quantum theory suggests the need to add higher-curvature terms. Moreover, string theoretic arguments~\cite{string lovelock} suggest that quadratic Lovelock gravity appears in the low-energy limit for strings propagating in curved spacetime.
 
For the above reasons, Lovelock gravity has been extensively investigated, with emphasis on the similarities to and difference from general relativity.
(See~\cite{lovelockreview} for review.)
However, in comparison with its classical aspects, the quantum theory is poorly understood at present, despite the fact that deep issues such as the origin of the four-dimensional universe can only be answered in the quantum context. 
In four-dimensional general relativity, Kucha\v{r} presented an elegant geometrical framework for studying the physical phase space and quantization of spherically symmetric vacuum black holes~\cite{kuchar94}. Using Kucha\v{r} geometrodynamics as a foundation, Louko and M\"akel\"a were able present a rigorous quantization of the Schwarzschild black hole spacetime, including a construction of all self-adjoint extensions of the Hamiltonian and derivation of the semi-classical area spectrum~\cite{Louko1996}. They found that the area/entropy spectrum was equally spaced in the semi-classical limit, in agreement with early speculations of Bekenstein
and Mukhanov based on the thermodynamic properties of black holes~\cite{Bekenstein}. An equally spaced area spectrum was also obtained for Schwarzschild black holes using a variety of different techniques~\cite{area spectrum}. 
This result is perhaps not surprising because there is only one length scale in the system, namely the Planck length.
In contrast, there is more than one length scale in quantum Lovelock gravity because the coupling constants to each order of the Lovelock Lagrangian are dimensionful. As a consequence, the entropy of Lovelock black holes is no longer equal to 1/4 the area~\cite{whitt1988}. It is therefore of great interest to see what quantum spectrum emerges for the both the area and entropy.

It is important to mention two key features of general relativity that were crucial to Kucha\v{r}'s analysis. First the existence of  Birkhoff's theorem implies that the reduced phase space is finite dimensional (two dimensional in the case of Schwarzschild black holes). Secondly, Kucha\v{r} identified the Misner-Sharp mass~\cite{ms1964} in spherically symmetric vacuum spacetime and its conjugate momentum, the Schwarzschild time separation at infinity, as the physical phase space variables. In more general situations (i.e. non-vacuum), the
Misner-Sharp mass is known as the best quasi-local mass in spherically symmetric spacetime. It satisfies the requisite monotonic and 
positivity properties and converges to the Arnowitt-Deser-Misner (ADM) mass at spacelike infinity in asymptotically flat spacetime~\cite{hayward1996}.
Fortunately, Lovelock gravity also possesses both these key features.
Bikrhoff's theorem in Lovelock gravity asserts that the spherically symmetric vacuum solution  is uniquely determined under certain conditions~\cite{zegers2005}.
The corresponding Schwarzschild-Tangherlini-type vacuum solution was obtained by Zegers~\cite{zegers2005}.
In addition, a natural counterpart to the Misner-Sharp mass has been defined in Lovelock gravity~\cite{mwr2011,HM08}.

The purpose of the present paper is to provide a framework to study quantum aspects of spherically symmetric black holes in Lovelock gravity.
 We  provide a comprehensive analysis that goes far beyond the initial presentation of our results in~\cite{GTMletter}.
In particular, we use the geometrodynamical formulation of Kucha\v{r} to do a complete Hamiltonian analysis, including derivation of the super-Hamiltonian and super-momentum constraints and verification of suitable boundary conditions for asymptotically flat black holes. Our analysis leads to a fully reduced Hamiltonian that is just as simple as that of Kucha\v{r}. As a specific application, we also derive the fully reduced equations of motion in flat slice coordinates for the collapse of a charged scalar field, including Lovelock gravitational as well as electromagnetic self-interactions. 

We note that the Hamiltonian analysis for full Lovelock gravity was first considered by Teitelboim and Zanelli~\cite{TZ}. Their result was rather formal in that an explicit parametrization of the phase space was not provided. (See also~\cite{df2012,st2008}.) For the case of spherical symmetry, the geometrodynamics~\cite{kuchar94} of five-dimensional Einstein-Gauss-Bonnet (i.e. quadratic Lovelock) gravity was worked out by Louko {\it et al}~\cite{JL97}, while the Hamiltonian analysis of higher-dimensional Gauss-Bonnet gravity coupled to matter was recently done in~\cite{TLKM}.
Our analysis was done for generic Lovelock gravity in arbitrary dimensions.

In the following section, we present our system, including action and spherically symmetric solutions. Section~\ref{Dim red action} derives a dimensionally reduced equivalent two-dimensional action that is the starting point of our analysis. It also reviews the geometrodynamics of Kucha\v{r} in a general dynamical setting. 
In Section~\ref{Can formalism gr}, we perform the Hamiltonian analysis for general relativity, in terms of both the standard ADM and geometrodynamical variables. The generalization to Lovelock gravity is presented in Section~\ref{Can form Lovelock}.
The contributions of matter fields are discussed in Section~\ref{matter}, while
concluding remarks and discussions appear in Section~\ref{Conclusions}. Detailed derivations and analysis of the boundary conditions are deferred to Appendices.

Our basic notation follows~\cite{wald}.
The convention for the Riemann curvature tensor is $[\nabla _\rho ,\nabla_\sigma]V^\mu ={{\cal R}^\mu }_{\nu\rho\sigma}V^\nu$ and ${\cal R}_{\mu \nu }={{\cal R}^\rho }_{\mu \rho \nu }$.
The Minkowski metric is taken as diag$(-,+,\cdots,+)$, and Greek indices run over all spacetime indices.
We adopt the units in which only the $n$-dimensional gravitational constant $G_n$ is retained.

\section{Preliminaries}
\label{Preliminaries}
\subsection{Symmetric spacetimes in Lovelock gravity}
The action of the gravitational system is written as
\begin{align}
I=I_{{\cal M}}+I_{\partial{\cal M}},\label{action}
\end{align}
where $I_{{\cal M}}$ is the dynamical term and $I_{\partial{\cal M}}$ is the boundary term.
In general relativity, $I_{{\cal M}}$ is the Einstein-Hilbert action and $I_{\partial{\cal M}}$ is the Gibbons-Hawking-York boundary term.
(See~\cite{olea2007} for the boundary term in general Lovelock gravity.)
In the present paper, we consider Lovelock gravity in $n (\geq 4)$-dimensional spacetime, of which the dynamical term in the action is given by
\begin{align}
\label{action2}
I_{{\cal M}}=&\frac{1}{2\kappa_n^2}\int \D ^nx\sqrt{-g}\sum_{p=0}^{[n/2]}\alpha_{(p)}{\ma L}_{(p)}+I_{\rm matter},\\
{\ma L}_{(p)}:=&\frac{1}{2^p}\delta^{\mu_1\cdots \mu_p\nu_1\cdots \nu_p}_{\rho_1\cdots \rho_p\sigma_1\cdots \sigma_p}{\cal R}_{\mu_1\nu_1}^{\phantom{\mu_1}\phantom{\nu_1}\rho_1\sigma_1}\cdots {\cal R}_{\mu_p\nu_p}^{\phantom{\mu_p}\phantom{\nu_p}\rho_p\sigma_p},
\end{align}
where $\kappa_n := \sqrt{8\pi G_n}$.
Our notation basically follows~\cite{mwr2011}.
$\alpha_{(p)}$ is the coupling constant for the $p$th-order Lovelock Lagrangian with dimension $({\rm length})^{2(p-1)}$ and we assume $\kappa_n^2>0$ without any loss of generality.
The $\delta$ symbol denotes a totally anti-symmetric product of Kronecker deltas, normalized to take values $0$ and $\pm 1$, defined by
\begin{align}
\delta^{\mu_1\cdots \mu_p}_{\rho_1\cdots \rho_p}:=&p!\delta^{\mu_1}_{[\rho_1}\cdots \delta^{\mu_p}_{\rho_p]}.
\end{align}

The gravitational equation following from this action is given by 
\begin{align} 
{\ma G}_{\mu\nu}=\kappa_n^2 {T}_{\mu\nu}, \label{beqL}
\end{align} 
where ${T}_{\mu\nu}$ is the energy-momentum tensor for matter fields obtained from $I_{\rm matter}$ and
\begin{align} 
{\ma G}_{\mu\nu} :=& \sum_{p=0}^{[n/2]}\alpha_{{(p)}}{G}^{(p)}_{\mu\nu}, \label{generalG}\\
{G}^{\mu(p)}_{~~\nu}:=& -\frac{1}{2^{p+1}}\delta^{\mu\eta_1\cdots \eta_p\zeta_1\cdots \zeta_p}_{\nu\rho_1\cdots \rho_p\sigma_1\cdots \sigma_p}{\cal R}_{\eta_1\zeta_1}^{\phantom{\eta_1}\phantom{\zeta_1}\rho_1\sigma_1}\cdots {\cal R}_{\eta_p\zeta_p}^{\phantom{\eta_p}\phantom{\zeta_p}\rho_p\sigma_p}.
\end{align} 
The tensor ${G}^{(p)}_{\mu\nu}$ obtained from ${\ma L}_{(p)}$ contains up to the second derivatives of the metric and ${G}^{(p)}_{\mu\nu}\equiv 0$ is satisfied for $p\ge [(n+1)/2]$.

In the present paper, we consider the $n(\ge 4)$-dimensional warped product spacetime $({\cal M}^n,g_{\mu\nu}) \approx ({M}^2,g_{AB})\times ({K}^{n-2},\gamma_{ab})$ with the general metric 
\begin{eqnarray}
g_{\mu\nu}(x)dx^\mu dx^\nu=g_{AB}({\bar y})d{\bar y}^A d{\bar y}^B+R({\bar y})^2\gamma_{ab}(z)dz^adz^b,
\label{eq:structure}
\end{eqnarray}
where $g_{AB}$ is an arbitrary Lorentz metric on $({M}^2,g_{AB})$ and $R({\bar y})$ is a scalar function on  $({M}^2,g_{AB})$.
$\gamma_{ab}$ is the metric on the $(n-2)$-dimensional maximally symmetric space $({K}^{n-2},\gamma_{ab})$ with its sectional curvature $k=1,0,-1$. 
We note that the results in the present paper are valid for $k=0$ with $p=0$ by setting $k^p=1$.
We introduce the covariant derivatives on spacetime $({\ma M}^n,g_{\mu\nu})$, the subspacetime $({M}^2,g_{AB})$ and the maximally symmetric space $({K}^{n-2},\gamma_{ab})$ with
\begin{eqnarray}
\nabla_\rho g_{\mu\nu}=0,\qquad D_F g_{AB}=0,\qquad {\bar D}_fg_{ab}=0.
\end{eqnarray}

The most general energy-momentum tensor $T_{\mu\nu}$ compatible with this spacetime symmetry governed by Lovelock equations is given by
\begin{align}
T_{\mu\nu}\D x^\mu \D x^\nu =T_{AB}({\bar y})\D {\bar y}^A\D {\bar y}^B+p({\bar y})R^2 \gamma_{ab}\D z^a\D z^b,
\end{align}  
where $T_{AB}({\bar y})$ and $p({\bar y})$ are a symmetric two-tensor and a scalar on $(M^2, g_{AB})$, respectively.

The generalized Misner-Sharp mass in Lovelock gravity is defined by
\begin{align}
M :=& \frac{(n-2)V_{n-2}^{(k)}}{2\kappa_n^2}\sum_{p=0}^{[n/2]}{\tilde \alpha}_{(p)}R^{n-1-2p}[k-(DR)^2]^p,\label{qlm-L}\\
{\tilde \alpha}_{(p)}:=&\frac{(n-3)!\alpha_{(p)}}{(n-1-2p)!}, \label{alphatil}
\end{align}  
where $(DR)^2:=(D_A R)(D^A R)$~\cite{mwr2011}.
The constant $V_{n-2}^{(k)}$ represents the volume of $({K}^{n-2},\gamma_{ab})$ if it is compact and otherwise arbitrary positive.
$M$ reduces to the ADM (Arnowitt-Deser-Misner) mass at spacelike infinity in the asymptotically flat spacetime.
In terms of $M$, some components of the Lovelock equation are written in the following simple form~\cite{mwr2011,tw2011}:
\begin{align}
D_A  M =&V_{n-2}^{(k)}R^{n-2}\biggl({T_A}^B(D_B R) -{T^B}_B (D_A R)\biggl). \label{1stlaw1}
\end{align}

\subsection{Vacuum solutions}
In the vacuum case ($T_{\mu\nu}=0$), Eq.~(\ref{1stlaw1}) shows that $M$ is constant.  
The maximally symmetric solution, namely Minkowski, de~Sitter (dS) or anti-de~Sitter (AdS) solution, gives $M=0$.
The maximally symmetric spacetime may be given in the following coordinates:
\begin{align}
ds^2=&-(k-{\tilde \lambda}r^2)dt^2+\frac{dr^2}{k-{\tilde \lambda}r^2}+r^2\gamma_{ab}dz^a dz^b,\label{vacuum} 
\end{align}
where ${\tilde \lambda}:=2\lambda/[(n-1)(n-2)]$ and $\lambda$ is the effective cosmological constant, which is determined by the following algebraic equation:
\begin{align}
\label{lambda}
0=\sum_{p=0}^{[n/2]}{\tilde \alpha}_{(p)}{\tilde \lambda}^p=:v({\tilde\lambda}).
\end{align}  
The Minkowski vacuum ($\lambda=0$) is possible only if ${\alpha}_{(0)}=0$.
Since Eq.~(\ref{lambda}) is a higher-order polynomial, there can be multiple values of ${\tilde \lambda}$.
We call the vacuum ${\tilde \lambda}={\tilde \lambda}_1$ {\it non-degenerate} if $(dv/d{\tilde\lambda})({\tilde \lambda}_1)\ne 0$ holds.
A {\it simply} degenerate vacuum is characterized by $(dv/d{\tilde\lambda})({\tilde \lambda}_1)= 0$, while a {\it doubly} degenerate vacuum is characterized by $(dv/d{\tilde\lambda})({\tilde \lambda}_1)= (d^2v/d{\tilde\lambda}^2)({\tilde \lambda}_1)= 0$.
In a similar manner, a $q$th-order degenerate vacuum is defined by $(d^sv/d{\tilde\lambda}^s)({\tilde \lambda}_1)=0$ for $s=1,2,\cdots,q$, where $q \le [(n-3)/2]$ is satisfied because of ${\tilde \alpha}_{(n/2)}\equiv 0$ for even $n$.

The Schwarzschild-Tangherlini-type vacuum solution in Lovelock gravity~\cite{zegers2005} is given by
\begin{align}
ds^2=-f(r)dt^2+\frac{dr^2}{f(r)}+r^2\gamma_{ab}dz^a dz^b, \label{f-vacuum}
\end{align}
where the metric function $f(r)$ is determined algebraically by 
\begin{align}
\label{alg}
{\tilde M} =\sum_{p=0}^{[n/2]}{\tilde \alpha}_{(p)}r^{n-1-2p}(k-f(r))^p.
\end{align}
${\tilde M}$ is related to the constant generalized Misner-Sharp mass as ${\tilde M}:=2\kappa_n^2M/[(n-2)V_{n-2}^{(k)}]$.
This class of vacuum solutions in higher-order Lovelock gravity was first obtained by Boulware and Deser~\cite{bdw} in the quadratic theory.
(See~\cite{bdw2} for further discussions.)
The above solution reduces to the ones found in~\cite{DCBH,DCBH2} in the case where the coupling constants are chosen such that the theory admits a fully degenerate maximally symmetric vacuum.
Birkhoff's theorem in Lovelock gravity asserts that, in the case where $(DR)^2:=(D_AR)(D^AR)\ne 0$ and the spacetime is of the $C^2$-class, there is a unique vacuum solution as long as the theory does not admit degenerate vacua~\cite{zegers2005}. (See also~\cite{mwr2011} for more general case.)
One of the purposes of the present paper is to derive the formulae to quantize a Lovelock black hole described by the above solution.
 
If the theory admits degenerate vacua, there may be more vacuum solutions.
If the theory admits simply degenerate vacua, the following is also a vacuum solution:
\begin{align}
ds^2=&-(k-{\tilde\lambda} r^2)e^{2\delta(t,r)}dt^2+\frac{dr^2}{k-{\tilde\lambda} r^2}+r^2\gamma_{ab}dz^a dz^b, \label{type-I} 
\end{align}  
where $\delta(t,r)$ is an {\it arbitrary} function and ${\tilde\lambda}$ takes the value for the degenerate vacuum.
(This solution was first obtained properly in quadratic Lovelock gravity by Charmousis and Dufaux~\cite{cd2002}.)
If the theory admits doubly degenerate vacua, there is another vacuum solution where the two-dimensional portion $(M^2, g_{AB})$ is {\it totally arbitrary}~\cite{mwr2011}.
{If one removes the assumption of  $C^2$-differentiability of the spacetime,  then more vacuum solutions exist~\cite{Garraffo:2007fi,Gravanis:2010zs}.}

\section{Dimensionally reduced action}
\label{Dim red action}
\subsection{Covariant form}
In the symmetric spacetime under consideration, the system may be described by the effective two-dimensional action: 
\begin{align}
I_{(2)}=I_{M}+I_{\partial {M}}.\label{action3}
\end{align}
The dynamical term $I_{M}$ is written as
\begin{align}
I_{M}=\int d{\bar y}^0L[g_{AB},R]=\int d{\bar y}^0 \int d{\bar y}^1{\cal L}[g_{AB},R],\label{2-action}
\end{align}
where ${\bar y}^0$ is a timelike coordinate on $(M^2, g_{AB})$.
Here the Lagrangian $L$ and the Lagrangian density ${\cal L}$ are functionals of the metric functions, which are determined up to a total derivative.
The main purpose of geometrodynamics is to find canonical variables (that are functionals of the metric functions) to provide a tractable form and transparent physical meaning for ${\cal L}$.

Our first task is to derive a tractable tensorial form of $I_{M}$.
For symmetric spacetimes under consideration, the action reduces to
\begin{align}
I_{M}=\frac{V_{n-2}^{(k)}}{2\kappa_n^2}\int d^2{\bar y}\sqrt{-g_{(2)}}R^{n-2}\sum^{[n/2]}_{p=0}\alpha_{(p)} {\cal L}_{(p)},
\label{eq:reduced action 2}
\end{align}
where $g_{(2)}:=\det(g_{AB})$ and the dimensionally reduced $p$th-order Lovelock term $ {\cal L}_{(p)}$ is given from expressions (2.19) and (2.20) of~\cite{mwr2011} as 
\begin{align}
{\cal L}_{(p)} =& \frac{(n-2)!}{(n-2p)!} \biggl[(n-2p)(n-2p-1)\left(\frac{k-(DR)^2}{R^2}\right)^{p} - 2p(n-2p)\frac{D^2R}{R}\left(\frac{k-(DR)^2}{R^2}\right)^{p-1} \nonumber \\
& + 2p(p-1) \frac{(D^2R)^2 - (D^AD_BR)(D^BD_AR)}{R^2} \left(\frac{k-(DR)^2}{R^2}\right)^{p-2} + p\overset{(2)}{\cal R} \left(\frac{k-(DR)^2}{R^2}\right)^{p-1} \biggl],  \label{L_p}
\end{align}
where ${}^{(2)}{\cal R}$ is the Ricci scalar on $(M^2, g_{AB})$ and $D^2R:=D^AD_AR$.
At a glance, there is a non-minimal coupling between $(DR)^2$ and ${}^{(2)}{\cal R}$ in ${\cal L}_{(p)}$.
Such a Lagrangian is not tractable to perform the canonical analysis.
As proven in Appendix~\ref{appendix1}, we can write it, up to total divergences, without such a coupling: 
\begin{align}
\label{eq:lovelock simplified}
 {\cal L}_{(p)} =& \frac{(n-2)!}{(n-2p)!} \Biggl[pk^{p-1}\overset{(2)}{\cal R} R^{2-2p} + pR^{2-n}\frac{D^A(R^{n-2p})D_A((DR)^2)}{(DR)^2}\biggl\{k^{p-1}-(k-(DR)^2)^{p-1}\biggl\}  \nonumber \\
& + (n-2p)(n-2p-1)\biggl\{\left(k-(DR)^2\right)^{p} +2pk^{p-1}(DR)^2\biggl\}R^{-2p} \Biggr]. 
\end{align}
This is a key result of our paper and the starting point of our canonical analysis.

Note that in the following we work exclusively with the equations of motion derived from the reduced action (\ref{eq:reduced action 2}). In general it is not true that dimensional reduction commutes with the variational principle. That is, the space of extrema of a dimensionally reduced action in principle may not coincide with the space of symmetric solutions of the unreduced action. However, in a very elegant and powerful set of papers \cite{Palais1979,Fels2002} (see also \cite{Deser2003}), it has been rigorously proven that if the symmetry group is a compact Lie group, as in our case, then for any local metric theory of gravity in arbitrary space-time dimensions, with or without matter, variation does indeed commute with dimensional reduction. The spherically symmetric equations of motion obtained from the full, unreduced Lovelock action with matter were explicitly written down in \cite{mwr2011}.  The proof that the solution space is the same in both cases nonetheless requires the more detailed analysis of \cite{Palais1979,Fels2002}. Unfortunately, this analysis is only valid in the compact case, so that more work needs to be done in order to prove that the dimensionally reduced action is sufficient when $k=0,-1$. This is one of the reasons that we defer consideration of the non-compact case to a future study.

\subsection{ADM form}
We are going to write down the action (\ref{eq:lovelock simplified}) by adopting the following ADM coordinates $(t,x)$ on $(M^2,g_{AB})$:
\begin{eqnarray}
ds_{(2)}^2=g_{AB}d{\bar y}^Ad{\bar y}^B=-N(t,x)^2dt^2+\Lambda(t,x)^2(dx+N_r(t,x)dt)^2.\label{ADM}
\end{eqnarray} 
Now canonical variables are $N$, $N_r$, $\Lambda$, and $R$ and their momentum conjugates are respectively written as $P_{N}$, $P_{N_r}$, $P_{\Lambda}$, and $P_{R}$.
In the present paper, a dot and a prime denote a partial derivative with respect to $t$ and $x$, respectively.
The metric and its inverse are
\begin{align}
g_{tt}=&-(N^2-\Lambda^2N_r^2),\quad g_{tx}=\Lambda^2N_r,\quad g_{xx}=\Lambda^2,\\
g^{tt}=&-N^{-2},\quad g^{tx}=N_rN^{-2},\quad g^{xx}=N^{-2}\Lambda^{-2}(N^2-\Lambda^2N_r^2),
\end{align} 
while $\sqrt{-g_{(2)}}$ is given by  
\begin{align}
\sqrt{-g_{(2)}}=N\Lambda.
\end{align} 
For the later use, we compute the following quantities:
\begin{align}
F:=&(DR)^2 \nonumber \\
=&-y^2+\Lambda^{-2}{R'}^2,\label{defF}\\
\sqrt{-g_{(2)}}D^2 R=&-\partial_t(\Lambda y)+\partial_x(\Lambda N_r y+\Lambda^{-1}NR'),
\end{align} 
where $y$ is defined by 
\be
y:=N^{-1}({\dot R}-N_rR').\label{defy}
\ee 

We also need the following relationship:
\bea
D^A(R^{n-2p})D_A((DR)^2)
  &=& 
  (n-2p) R^{n-2p-1} \left(  - \frac{1}{N}y \dot{F}
+ \left(\frac{R'}{\Lambda^2}
   +\frac{N_r}{N} y\right) F'\right).
\eea
Using this result, the action (\ref{eq:lovelock simplified}) is written in the following simple form:
\begin{align}
I_M=&\frac{(n-2)V_{n-2}^{(k)}}{2\kappa_n^2}\sum^{[n/2]}_{p=0}\int d^2{\bar y}\sqrt{-g_{(2)}}\frac{{\tilde \alpha}_{(p)} }{(n-2p)} \Biggl[pk^{p-1}\overset{(2)}{\cal R}R^{n-2p} \nonumber \\
&- p(n-2p)\frac{R^{n-2p-1}}{N\Lambda}\{k^{p-1}-(k-F)^{p-1}\}\biggl\{\Lambda y\frac{\dot F}{F}-(\Lambda N_r y+\Lambda^{-1}NR')\frac{F'}{F}\biggl\} \nonumber \\
& + (n-2p)(n-2p-1)\biggl\{\left(k-F\right)^{p} +2pk^{p-1}F\biggl\}R^{n-2-2p} \Biggr]. \label{action-3}
\end{align}
The first term is the two-dimensional gravity non-minimally coupled scalar field $R$, which is essentially the same as the general relativistic case.
This term can be explicitly written down in terms of the canonical variables using
\begin{align}
\sqrt{-g_{(2)}}R^{n-2p}\overset{(2)}{\cal R}=&-2N^{-1}\biggl((R^{n-2p})'N_r-\partial_t(R^{n-2p})\biggl)(N_r'\Lambda +N_r\Lambda'-{\dot \Lambda}) \nonumber \\
&-2N\biggl((R^{n-2p})''\Lambda^{-1}+(R^{n-2p})'(\Lambda^{-1})'\biggl)+\partial_t(\cdots)+\partial_x (\cdots).\label{intR}
\end{align} 
Based on the action (\ref{action-3}), we will perform the canonical analysis in the subsequent sections using geometrodynamical phase space variables. We therefore now review briefly the geometrodynamics of Kucha\v{r}~\cite{kuchar94}.

\subsection{Geometrodynamics}
The metric (\ref{ADM}) may be written in the generalized Schwarzschild form in terms of the areal coordinates as
\begin{align}
ds_{(2)}^2=-F(R,T)e^{2\sigma(R,T)}dT^2+F(R,T)^{-1}dR^2.
\label{eq:Schwarzschild metric}
\end{align}  
The generalized Misner-Sharp mass $M$ is then given by
\begin{align}
M(R,T)= \frac{(n-2)V_{n-2}^{(k)}}{2\kappa_n^2}\sum_{p=0}^{[n/2]}{\tilde \alpha}_{(p)}R^{n-1-2p}\biggl(k-F(R,T)\biggl)^p.\label{M-F}
\end{align}  
This implicitly gives the functional form $F=F(R,M)$.
However, there is no one-to-one correspondence between $F$ and $M$ unless all the coupling constants $\alpha_{(p)}$ are non-negative.

To see the relation to the ADM form (\ref{ADM}) we use the coordinate transformations $T=T(t,x)$ and $R=R(t,x)$, to write the metric (\ref{eq:Schwarzschild metric}) as
\begin{align}
ds_{(2)}^2=&-(F{\dot T}^2e^{2\sigma}-F^{-1}{\dot R}^2)dt^2+2(-F{\dot T}T'e^{2\sigma}+F^{-1}{\dot R}R')dtdx \nonumber \\
&+(-F{T'}^2e^{2\sigma}+F^{-1}{R'}^2)dx^2.
\end{align}  
Comparing with the ADM form, we identify
\begin{align}
F{\dot T}^2e^{2\sigma}-F^{-1}{\dot R}^2=&N^2-\Lambda^2{N_r}^2,\\
-F{\dot T}T'e^{2\sigma}+F^{-1}{\dot R}R'=&\Lambda^2N_r,\\
-F{T'}^2e^{2\sigma}+F^{-1}{R'}^2=&\Lambda^2 \label{Gamma1}
\end{align} 
and obtain
\begin{align}
N_r=&\frac{-F{\dot T}T'e^{2\sigma}+F^{-1}{\dot R}R'}{-F{T'}^2e^{2\sigma}+F^{-1}{R'}^2},\label{beta}\\
N=& \frac{e^{\sigma}({\dot T}R'-{\dot R}T')}{\sqrt{-F{T'}^2e^{2\sigma}+F^{-1}{R'}^2}},\label{alpha} 
\end{align} 
{ As discussed by Kucha\v{r} in Section IVA of~\cite{kuchar94}, one can ensure that ${\dot T}R'-{\dot R}T'$ and hence the Lapse function $N$ are positive by an appropriate choice of $x$.}
$y$ is then given from the definition (\ref{defy}) as 
\begin{align}
y=\frac{FT'e^{\sigma}}{\sqrt{-F{T'}^2e^{2\sigma}+F^{-1}{R'}^2}},
\end{align} 
from which we obtain
\begin{align}
T'e^{\sigma}=\frac{y\Lambda}{F},\label{T-sigma}
\end{align} 
where we used Eq.~(\ref{Gamma1}).
Using this to eliminate $T'e^{\sigma}$ in Eq.~(\ref{Gamma1}), we obtain
\begin{align}
F=-y^2+\frac{{R'}^2}{\Lambda^{2}}
\label{eq:F1}
\end{align} 
as required by consistency with (\ref{defF})
 
In the above, we derived expressions for the generalized Schwarzschild time $T$ in terms of the canonical ADM variables. As we will see in the following this determines the conjugate momentum to the Misner-Sharp mass function in a form that is appropriate for slicings that approach the Schwarzschild form at spatial infinity. Other asymptotic forms for the slicings are possible, including flat slice or generalized Painlev\'{e}-Gullstrand (PG) coordinates:
\be
ds_{(2)}^2=-{e}^{2\sigma}dT_{\rm PG}^2 + (dR + G{e}^{\sigma} dT_{\rm PG})^2,
\label{p-g-metric}
\ee
where $\sigma=\sigma(T_{\rm PG},R)$ and $G=G(T_{\rm PG},R)$.
The geometrodynamical variables appropriate for such slicings  were first derived in \cite{Louko2007}.
Since we have 
\be
(D R)^2=1-G^2
\ee 
for the above form of the metric, it follows that
\be
G=\pm\sqrt{1-F}.
\ee
By inspection of (\ref{p-g-metric}) one can see that the positive sign yields an equation for ingoing null geodesics that is regular at any horizon $F=0$, so this is the choice that is suitable for describing the spacetime near a future horizon (black hole). The opposite sign must be chosen for a past horizon (white hole).
We now go through exactly the same derivation as before.
{Performing the coordinate transformations $T_{\rm PG}=T_{\rm PG}(t,x)$ and $R=R(t,x)$ in the metric (\ref{p-g-metric}) and comparing to the ADM form (\ref{ADM}) yields: }
\begin{subequations}%
\label{eq:rec1}%
\bea
\Lambda^2
&=& 
(R'+{e}^{\sigma} G T_{\rm PG}')^2 -{e}^{2\sigma}{T_{\rm PG}'}^2, 
\label{conformal-factor}
\\
N^2-\Lambda^2N_r^2
&=&{e}^{2\sigma} \dot{T}_{\rm PG}^2
- (\dot{R}+{e}^{\sigma} G \dot{T}_{\rm PG})^2, 
\label{eq:mixed}
\\
\Lambda^2 N_r
&=& 
(R'+{e}^{\sigma} G T_{\rm PG}')(\dot{R}+{e}^{\sigma} G \dot{T}_{\rm PG})  -{e}^{2\sigma}T_{\rm PG}'\dot{T}_{\rm PG}. 
\label{shift}
\eea
\end{subequations}
Solving (\ref{eq:rec1}) for $N$ and~$N_r$, we find%
\begin{subequations}%
\label{eq:N-and-sigma-solved}%
\bea
N_r 
&=& \frac{(R'+{e}^{\sigma} G T_{\rm PG}')(\dot{R}+{e}^{\sigma} G \dot{T}_{\rm PG}) -{e}^{2\sigma}T_{\rm PG}'\dot{T}_{\rm PG}}{(R'+{e}^{\sigma} G T_{\rm PG}')^2-{{e}^{\sigma}(T_{\rm PG}')}^2} ,
\\
N 
&=& \frac{R'{e}^{\sigma}\dot{T}_{\rm PG}- \dot{R}{e}^{\sigma} T_{\rm PG}'}{\sqrt{(R'+{e}^{\sigma} G T_{\rm PG}')^2-{{e}^{\sigma}(T_{\rm PG}')}^2}} . 
\label{eq:sigma-solved}%
\eea
\end{subequations}
To complete the derivation, we use (\ref{defy}) and the above expressions for $\Lambda$, $N$ and $N_r$ to calculate
\bea
y\Lambda &=& (1-G^2)e^{\sigma}T_{\rm PG}' + G R',
\label{eq:yLambda PG}
\eea
which yields:
\be
e^{\sigma}T_{\rm PG}' = \frac{y \Lambda}{F} \pm \frac{R'\sqrt{1-F}}{F}.
\label{eq:Tprime PG}
\ee
The second term on the right-hand side of the above guarantees that the PG time is well defined either for (with a +ve sign) future  or (with a -ve sign) past  horizons. 

\section{Canonical formalism in general relativity}
\label{Can formalism gr}
In this section, we perform the canonical analysis for spherically symmetric spacetimes ($k=1$) in general relativity without a cosmological constant, which is a generalization of the Kucha\v{r}'s analysis in four dimensions to arbitrary dimensions.
We set ${\tilde\alpha}_{(1)}=1$ in this section for simplicity.
The reduced action (\ref{action-3}) then becomes quite simple:
\begin{align}
I_{\rm M(GR)}=&\frac{\ma A_{n-2}}{2\kappa_n^2}\int d^2{\bar y}  \Biggl[2(n-2)R^{n-3}y(N_r'\Lambda +N_r\Lambda') -2N\biggl((R^{n-2})''\Lambda^{-1}+(R^{n-2})'(\Lambda^{-1})'\biggl)  \nonumber \\
& + (n-2)(n-3)(1+F)N\Lambda R^{n-4}-2(n-2)R^{n-3}y {\dot \Lambda}\Biggr]. \label{action-3gr}
\end{align}
Here $\ma A_{n-2}$ is the surface area of an $(n-2)$-dimensional unit sphere, namely
\begin{align}
\ma A_{n-2} :=\frac{2\pi^{(n-1)/2}}{\Gamma((n-1)/2)}(\equiv V_{n-2}^{(1)}),
\label{unitarea}
\end{align}
where $\Gamma(x)$ is the Gamma function.
The purpose of this section is to show that the areal radius and the Misner-Sharp mass are well-defined canonical variables in the system, which will be generalized to Lovelock gravity in the following 
.

\subsection{ADM variables}
We first derive the expressions for $P_\Lambda$ and $P_R$.
The corresponding Lagrangian density of the action (\ref{action-3gr}) is 
\begin{align}
{\cal L}=& \frac{(n-2)\ma A_{n-2}}{2\kappa_n^2} \Biggl[2R^{n-3}y(N_r'\Lambda +N_r\Lambda')-\frac{2}{(n-2)}N\biggl((R^{n-2})''\Lambda^{-1}+(R^{n-2})'(\Lambda^{-1})'\biggl)    \nonumber \\
&+ (n-3)(1+F)N\Lambda R^{n-4}-2R^{n-3}y {\dot \Lambda}\biggl],
\end{align}
from which we obtain $P_N=P_{N_r}=0$ and 
\begin{align}
P_\Lambda=& -\frac{(n-2)\ma A_{n-2}}{\kappa_n^2}R^{n-3}N^{-1}({\dot R}-N_rR'),\label{PLambdaGR2}\\
P_R=& \frac{(n-2)\ma A_{n-2} }{2\kappa_n^2} \Biggl[2R^{n-3}N^{-1}(N_r'\Lambda +N_r\Lambda')-2 (n-3)N^{-1}\Lambda R^{n-4}({\dot R}-N_rR')-2R^{n-3}N^{-1} {\dot \Lambda}\biggl].\label{PR}
\end{align}
With $P_\Lambda$ and $P_R$, the Hamiltonian density ${\cal H}(:={\dot \Lambda}P_{\Lambda}+{\dot R}P_{R}-{\cal L})$ is given by
\begin{align}
{\cal H}=&(N_r\Lambda P_{\Lambda})'-N_r\Lambda P_{\Lambda}'+N_rR'P_{R}-\frac{\kappa_n^2 N}{(n-2)\ma A_{n-2}R^{n-2}}P_{\Lambda}\biggl(RP_{R}-\frac{n-3}{2}\Lambda P_{\Lambda}\biggl)  \nonumber \\
&-\frac{(n-2)\ma A_{n-2}}{\kappa_n^2}N\biggl\{R^{n-3}\biggl(-R'' \Lambda^{-1}+R'\Lambda^{-2}\Lambda'\biggl)+\frac{n-3}{2} \Lambda R^{n-4}\biggl(1-\Lambda^{-2}{R'}^2\biggl)\biggl\}. \label{hamildens-gr}
\end{align} 
The first term in Eq.~(\ref{hamildens-gr}) is the total derivative and becomes a boundary term.
Since $P_N=P_{N_r}=0$, the Hamilton equations for $N$ and $N_r$ give constraint equations $H=0$ and $H_r=0$, where the super-momentum $H_{r}(:=\delta {\cal H}/\delta N_r)$ and the super-Hamiltonian $H(:=\delta {\cal H}/\delta N)$ are given by 
\begin{align}
H_r=&-\Lambda P_{\Lambda}'+R'P_{R}, \label{Hr}\\
H=&-\frac{\kappa_n^2}{(n-2)\ma A_{n-2}R^{n-2}}P_{\Lambda}\biggl(RP_{R}-\frac{n-3}{2}\Lambda P_{\Lambda}\biggl)  \nonumber \\
&-\frac{(n-2)\ma A_{n-2}}{\kappa_n^2}\biggl\{R^{n-3}\biggl(-R'' \Lambda^{-1}+R'\Lambda^{-2}\Lambda'\biggl)+\frac{n-3}{2} \Lambda R^{n-4}\biggl(1-\Lambda^{-2}{R'}^2\biggl)\biggl\}. \label{H}
\end{align} 
The action is finally written as
\begin{align}
I_{\rm M(GR)}=&\int dt\int dx ( {\dot \Lambda}P_{\Lambda}+{\dot R}P_{R}-N H-N_r H_{r}).\label{IMGR}
\end{align} 

It can be verified that with suitable boundary conditions the constraints $H$ and $H_r$ are first class in the Dirac sense and generate spacetime diffeomorphisms that preserve the spherically symmetric form of the metric.

\subsection{The Schwarzschild-Tangherlini spacetime in various coordinate systems}
In this subsection, we review various coordinate systems in the Schwarzschild-Tangherlini spacetime.
The Schwarzschild-Tangherlini vacuum solution in the best-known Schwarzschild coordinates is given by
\begin{align}
ds^2=&-\biggl(1-\frac{{\tilde M}}{r^{n-3}}\biggl)dt^2+\biggl(1-\frac{{\tilde M}}{r^{n-3}}\biggl)^{-1}dr^2+r^2\gamma_{ab}dz^a dz^b, \label{st} \end{align}
where ${\tilde M}:=2\kappa_n^2M/[(n-2){\cal A}_{n-2}]$ and $M$ is the ADM mass.
The corresponding ADM variables are
\begin{align}
N^2=1-\frac{{\tilde M}}{r^{n-3}},\quad N_r=0,\quad \Lambda^2=&\biggl(1-\frac{{\tilde M}}{r^{n-3}}\biggl)^{-1}, \quad R=r.
\end{align} 
In the next subsection, we will consider the boundary condition at spacelike infinity with this slicing.
However, there is a variety of slicings in the Schwarzschild-Tangherlini spacetime, as presented below.
 
By introducing a new spacelike coordinate $\rho$ as $r=\rho[1+{\tilde M}/(4\rho^{n-3})]^{2/(n-3)}$, the metric~(\ref{st}) is transformed into the isotropic coordinates: 
\begin{align}
ds^2=&-\frac{[1-{\tilde M}/(4\rho^{n-3})]^2}{[1+{\tilde M}/(4\rho^{n-3})]^2}dt^2+\left(1+\frac{{\tilde M}}{4\rho^{n-3}}\right)^{4/(n-3)}(d\rho^2+\rho^2\gamma_{ab}dz^a dz^b). \label{Sch-iso}
\end{align}
The corresponding ADM variables are
\begin{align}
N^2=&\frac{[1-{\tilde M}/(4\rho^{n-3})]^2}{[1+{\tilde M}/(4\rho^{n-3})]^2},\quad N_r=0,\quad \Lambda^2=\left(1+\frac{{\tilde M}}{4\rho^{n-3}}\right)^{4/(n-3)},\quad R=\rho\left(1+\frac{{\tilde M}}{4\rho^{n-3}}\right)^{2/(n-3)}.
\end{align} 
On the other hand, by introducing a new time coordinate $\tau$ defined by
\begin{equation}
d\tau:=dt+\sqrt{\frac{{\tilde M}}{r^{n-3}}}\frac{dr}{(1-{\tilde M}/r^{n-3})},
\end{equation}
the metric~(\ref{st}) is transformed into the Painlev\'{e}-Gullstrand coordinates:
\begin{equation}
ds^2=-\biggl(1-\frac{{\tilde M}}{r^{n-3}}\biggl)d\tau^2+2\sqrt{\frac{{\tilde M}}{r^{n-3}}}d\tau dr+dr^2+r^2\gamma_{ab}dz^a dz^b. \label{Sch-P}
\end{equation}
The corresponding ADM variables are
\begin{align}
N^2=1,\quad N_r=\sqrt{\frac{{\tilde M}}{r^{n-3}}},\quad \Lambda^2=1,\quad R=r.
\end{align}

By the coordinate transformation
\begin{equation}
r=\biggl(\frac{n-1}{2}\biggl)^{4/(n-1)}{\tilde M}^{1/(n-1)}\biggl(\frac{2({\tilde\rho}-\tau)}{n-1}\biggl)^{2/(n-1)}
\end{equation}
from the Painlev\'{e}-Gullstrand coordinates (\ref{Sch-P}), we obtain the Schwarzschild-Tangherlini metric in the Lema\^{\i}tre coordinates:
\begin{align}
ds^2=&-d\tau^2+{\tilde M}^{2/(n-1)}\biggl[\biggl(\frac{n-1}{2}({\tilde\rho}-\tau)\biggl)^{-2(n-3)/(n-1)}d{\tilde\rho}^2+\biggl(\frac{n-1}{2}({\tilde\rho}-\tau)\biggl)^{4/(n-1)}\gamma_{ab}dz^a dz^b\biggl]. \label{Sch-L}
\end{align}
In this coordinate system, the central curvature singularity and the black-hole event horizon are represented by $\tau={\tilde\rho}$ and 
\begin{align}
{\tilde\rho}-\tau=\frac{2{\tilde M}^{1/(n-3)}}{n-1},
\end{align}
respectively.
In the Lema\^{\i}tre coordinates, the radial coordinate ${\tilde\rho}$ does not coincide with $R$ in the asymptotically flat region.
Defining a new radial coordinate as
\begin{align}
{\tilde r}:={\tilde M}^{1/(n-1)}\biggl(\frac{n-1}{2}{\tilde\rho}\biggl)^{2/(n-1)},
\end{align}
which coincides with $R$ in the asymptotic region, we transform the metric (\ref{Sch-L}) into the following form:
\begin{align}
ds^2=&-d\tau^2+\biggl(1-\frac{(n-1){\tilde M}^{1/2}\tau}{2{\tilde r}^{(n-1)/2}}\biggl)^{-2(n-3)/(n-1)}d{\tilde r}^2
   \nonumber\\
 & +{\tilde r}^2\biggl(1-\frac{(n-1){\tilde M}^{1/2}\tau}{2{\tilde r}^{(n-1)/2}}\biggl)^{4/(n-1)}\gamma_{ab}dz^a dz^b. \label{Sch-L2}
\end{align}
This metric is inhomogeneous and time-dependent in this coordinate system and the corresponding ADM variables and their fall-off rates are
\begin{align}
N^2=&1,\quad N_r=0,\nonumber\\
\Lambda^2=&\biggl(1-\frac{(n-1){\tilde M}^{1/2}\tau}{2{\tilde r}^{(n-1)/2}}\biggl)^{-2(n-3)/(n-1)},\quad R={\tilde r}\biggl(1-\frac{(n-1){\tilde M}^{1/2}\tau}{2{\tilde r}^{(n-1)/2}}\biggl)^{2/(n-1)}.
\end{align}

Lastly, we present the Schwarzschild-Tangherlini metric in the Kerr-Schild coordinates:
\begin{align}
ds^2=&-d{\hat t}^2+dr^2+r^2\gamma_{ab}dz^a dz^b+\frac{{\tilde M}}{r^{n-3}}(d{\hat t}+dr)^2 \label{Sch-SD}\\
=&-\biggl(1-\frac{{\tilde M}}{r^{n-3}}\biggl)d{\hat t}^2+\frac{2{\tilde M}}{r^{n-3}}d{\hat t}dr+\biggl(1+\frac{{\tilde M}}{r^{n-3}}\biggl)dr^2+r^2\gamma_{ab}dz^a dz^b. \label{Sc-6}
\end{align}
The corresponding ADM variables are
\begin{align}
N^2=&\biggl(1+\frac{{\tilde M}}{r^{n-3}}\biggl)^{-1},\quad N_r=\frac{{\tilde M}}{r^{n-3}}\biggl(1+\frac{{\tilde M}}{r^{n-3}}\biggl)^{-1},\quad \Lambda^2=1+\frac{{\tilde M}}{r^{n-3}},\quad R=r.
\end{align}

\subsection{Boundary condition and boundary terms}
To perform the geometrodynamics, the boundary condition plays a crucial role.
In the present paper, we adopt the following boundary condition at spacelike infinity $x\to \pm \infty$\footnote{These boundary conditions are suited to asymptotically Schwarzschild slicings. The analogous boundary conditions for PG coordinates are given in \cite{Husain} and discussed in Appendix~\ref{app:boundary}.}:
\begin{align}
N\simeq& N_\infty(t)+\mathcal{O}(x^{-\epsilon_1}),\label{bc1}\\
N_r\simeq& N_r^\infty(t) x^{-(n-3)/2-\epsilon_2},\label{bc2}\\
\Lambda \simeq& 1+\Lambda_1(t)x^{-(n-3)},\label{bc3}\\
R\simeq& x+R_1(t)x^{-(n-4)-\epsilon_4},\label{bc4}
\end{align}
where $\epsilon_1$ is a positive number and $\epsilon_2$ and $\epsilon_4$ satisfy $\epsilon_2> \max[0,-(n-5)/2]$ and $\epsilon_4>\max[0,-(n-5)]$.
{ (The validity of this boundary condition is verified in Appendix~\ref{app:boundary}.)}
The asymptotic behavior of $P_\Lambda$ and $P_R$ are given by 
\begin{align}
P_\Lambda\simeq& -\frac{(n-2)\ma A_{n-2}}{\kappa_n^2}N_\infty^{-1}\biggl({\dot R_1} x^{1-\epsilon_4}-N_r^{\infty} x^{(n-3)/2-\epsilon_2}\biggl),\label{bc5}\\
P_R\simeq & -\frac{(n-2)\ma A_{n-2} }{\kappa_n^2} N_\infty^{-1}\biggl[N_r^{\infty}(t) \biggl(-\frac{n-3}{2}+\epsilon_2\biggl)x^{(n-5)/2-\epsilon_2}+{\dot \Lambda}_1(t)\biggl].\label{bc6}
\end{align}
Under the boundary condition adopted, the Misner-Sharp mass converges to a finite value $M\simeq M^\infty(t)$, where $M^\infty(t)$ is related to $\Lambda_1(t)$ as
\begin{align}
\Lambda_1(t)\equiv \frac{\kappa_n^2M^\infty(t)}{(n-2)\ma A_{n-2}}.\label{bc7}
\end{align} 

Now let us consider the boundary term for the action (\ref{IMGR}).
The role of the boundary term is to subtract the diverging terms at the boundary in the variation of the above action.
The action is completed by adding the boundary term, which gives a finite value in the variation.

Since the variation of $I_{\rm M(GR)}$ gives
\begin{align}
\delta I_{\rm M(GR)}=&\int dt\int dx \biggl( \partial_t (\delta {\Lambda}P_{\Lambda})-\delta \Lambda{\dot P_{\Lambda}}+ {\dot \Lambda}\delta P_{\Lambda}+ \partial_t (\delta {R}P_{R})-\delta R{\dot P_{R}}+{\dot R}\delta P_{R} \nonumber \\
&-\delta N H-N \delta H-\delta N_r H_{r}-N_r \delta H_{r}\biggl),\label{variation1gr}
\end{align} 
we need to know the contributions from $N_r\delta H_r$ and $N\delta H$.
Using the following results;
\begin{align}
N_r\delta H_r=&-N_r\delta \Lambda P_{\Lambda}'-(N_r\Lambda\delta P_\Lambda)'+(N_r\Lambda)' \delta P_{\Lambda}+(N_r\delta RP_R)'-\delta R(N_rP_{R})'+N_rR'\delta P_{R},\\
N\delta H=&\biggl(\mbox{irrelevant terms}\biggl)-\frac{(n-2)\ma A_{n-2}}{\kappa_n^2}\biggl\{-\biggl((NR^{n-3}\Lambda^{-1}\delta R)'-(NR^{n-3}\Lambda^{-1})'\delta R\biggl)' \nonumber \\
&+(N'R^{n-3}\Lambda^{-1}\delta R)'-(N'R^{n-3}\Lambda^{-1})'\delta R +(NR^{n-3}R'\Lambda^{-2}\delta \Lambda)'-(NR^{n-3}R'\Lambda^{-2})'\delta \Lambda\biggl\},
\end{align} 
we can write (\ref{variation1gr}) in the following form:
\begin{align}
\delta I_{\rm M(GR)}=&\int dt\int dx \biggl(\mbox{dynamical terms}\biggl)+\int dx \biggl[\delta {\Lambda}P_{\Lambda}+\delta {R}P_{R}\biggl]_{t=t_1}^{t=t_2} \nonumber \\
&-\int dt\biggl[-N_r\Lambda\delta P_\Lambda+N_rP_R\delta R \nonumber \\
&-\frac{(n-2)\ma A_{n-2}}{\kappa_n^2}\biggl\{-NR^{n-3}\Lambda^{-1}\delta (R')+N'R^{n-3}\Lambda^{-1}\delta R+NR^{n-3}R'\Lambda^{-2}\delta \Lambda\biggl\}\biggl]_{x=-\infty}^{x=+\infty}.
\end{align} 
Now the boundary condition comes into play.
We assume $\delta \Lambda=\delta R=0$ at $t=t_1,t_2$ and then the second term in the above variation vanishes. 
Using the boundary condition (\ref{bc1})--(\ref{bc6}), we can show that only the contribution in the last integral comes from $NR^{n-3}R'\Lambda^{-2}\delta \Lambda$ as
\begin{align}
NR^{n-3}R'\Lambda^{-2}\delta \Lambda\simeq N_\infty\delta \Lambda_1= \frac{\kappa_n^2N_\infty\delta M^\infty(t)}{(n-2)\ma A_{n-2}},
\end{align} 
where we used Eq.~(\ref{bc7}).
Finally we obtain the boundary term in a simple form:
\begin{align}
\delta I_{\rm M(GR)}=&\int dt\int dx \biggl(\mbox{dynamical terms}\biggl)+\int dt\biggl[N_\infty(t)\delta M^\infty(t)\biggl]_{x=-\infty}^{x=+\infty}. \label{boundaryGR}
\end{align}

\subsection{Misner-Sharp mass as canonical variable}
In the ADM coordinates, the canonical variables are $\{\Lambda, P_\Lambda; R,P_R\}$.
However, the physical meanings of the variable $\Lambda$ is not so clear.
In this subsection, we show that the two-dimensional equivalent action is written in a rather elegant manner by introducing the Misner-Sharp mass $M$ as a canonical variable.
We introduce a new set of canonical variables $\{M, P_M; S,P_S\}$ defined by
\begin{align}
S:=&R,
\label{eq:R}\\
P_S:=&P_R-\frac{1}{R'}(\Lambda P_\Lambda'+P_MM'),
\label{eq:PM}\\
M :=& \frac{(n-2)\ma A_{n-2}}{2\kappa_n^2}R^{n-3}(1-F),\label{MS}\\
P_M:=&-T'e^{\sigma}=-\frac{y\Lambda}{F},
\label{eq:Pr1}
\end{align}
where we used Eq.~(\ref{T-sigma}).
We are going to show below that, under the boundary condition (\ref{bc1})--(\ref{bc6}), the transformation from a set of variables $\{\Lambda, P_\Lambda; R, P_R\}$ to another set $\{M, P_M; S, P_S\}$ is a well-defined canonical transformation.

Note (\ref{eq:Pr1}) chooses the conjugate to $M$ in terms of the Schwarzschild time $T$. As verified in Appendix~\ref{app:boundary} this leads to a finite Liouville form providing one chooses boundary conditions such that the metric approaches the vacuum Schwarzschild solution sufficiently rapidly at spatial infinity. In order to use asymptotically PG slices, it is necessary to choose the conjugate to $M$ in terms of the PG time $T_{\rm PG}$. That is 
\be
\tilde{P}_M = -e^\sigma T'_{\rm PG}= \frac{y\Lambda}{F} - \frac{\sqrt{1-F}}{FR'}.
\label{eq:PM PG}
\ee
Since the extra term on the right is just a function of $R$ and $M$, this corresponds to a straightforward canonical transformation
$(M,P_M,S,P_S)\to (M,\tilde{P}_M,S,\tilde{P}_S)$.  
For simplicity, we henceforth stick to the Schwarzschild expressions.

From the expression (\ref{MS}) for the Misner-Sharp mass, we obtain
\begin{align}
P_M{\dot M} =&\frac{(n-2)\ma A_{n-2}}{2\kappa_n^2}\frac{y\Lambda}{F}R^{n-3}\biggl[{\dot F}-(n-3)(1-F)\frac{\dot R}{R}\biggl],\label{P_MdotMgr}
\end{align}  
which shows $P_M{\dot M} =P_\Lambda{\dot \Lambda}+(\cdots){\dot R}+\delta(\cdots)+(\cdots)'$.
The Misner-Sharp mass $M$ is expressed in terms of $\{\Lambda,P_\Lambda; R,P_R\}$ as
\begin{align}
M=&\frac{\kappa_n^2P_\Lambda^2}{2(n-2)\ma A_{n-2}R^{n-3}}+\frac{(n-2)\ma A_{n-2}}{2\kappa_n^2}R^{n-3}\biggl(1-\frac{{R'}^2}{\Lambda^2}\biggl).\label{MLiou}
\end{align}  
From this expression, we can show that $M'$ is a linear combination of the constraints:
\begin{align}
M'=\Lambda^{-1}(yH_r-R'H),\label{M'2}
\end{align}  
where we used Eq.~(\ref{PLambdaGR2}) to replace $P_\Lambda$ by $y$.
This implies that in the vacuum theory $M$ is a constant on the constraint surface, as expected.

Also using the expression (\ref{MLiou}), we can show that two sets of variables $\{\Lambda, P_\Lambda; R,P_R\}$ and $\{M, P_M; S,P_S\}$ satisfy the following Liouville form:
\begin{align}
P_\Lambda\delta \Lambda+P_R\delta R=P_M\delta M+P_S\delta S+\delta \eta+\zeta',\label{Liouville}
\end{align}
where
\begin{align}
\eta:=&\Lambda P_\Lambda+\frac{(n-2)\ma A_{n-2}}{2\kappa_n^2}R^{n-3}R' \ln\biggl|\frac{R'+y\Lambda}{R'-y\Lambda}\biggl|,\label{deta-gr}\\ 
\zeta:=&-\frac{(n-2)\ma A_{n-2}}{2\kappa_n^2}R^{n-3}\ln\biggl|\frac{R'+y\Lambda}{R'-y\Lambda}\biggl|\delta R. \label{zeta'-gr}
\end{align} 
Under the boundary condition (\ref{bc1})--(\ref{bc6}), the total derivative term $\zeta$ converges to zero at spacelike infinity.
Hence, the transformation from a set $\{\Lambda, P_\Lambda; R,P_R\}$ to $\{M,P_M;S,P_S\}$ is indeed a canonical transformation, namely
\begin{align}
&\int_{-\infty}^{\infty} dx(P_\Lambda\delta \Lambda+P_R\delta R)-\int_{-\infty}^{\infty} dx(P_M\delta M+P_S\delta S)=\delta \omega[\Lambda, P_\Lambda,; R,P_R], \label{canonical}\\
&\omega[\Lambda, P_\Lambda,; R,P_R]:=\int_{-\infty}^{\infty}dx\eta[\Lambda, P_\Lambda,; R,P_R]
\end{align} 
is satisfied.
It is shown that the integrands in the above equation, namely $P_\Lambda\delta \Lambda+P_R\delta R$, $P_M\delta M+P_S\delta S$, and $\eta$ converge to zero faster than $O(x^{-1})$ at spacelike infinity under the boundary condition we adopt, and hence the above expression is well-defined.
(See Appendix~\ref{app:boundary} for the proof.)

We now derive the Hamiltonian constraint and the diffeomorphism (momentum) constraint in terms of the variables $\{M,P_M;S,P_S\}$.
A straightforward calculation using the above equations verifies the following relation;
\begin{align}
{\cal L}-P_M{\dot M}-\frac{N\Lambda}{R'}M'=&-\frac{{\dot R}}{R'}P_MM'+\mbox{(t.d.)} \\
=&\frac{y\Lambda}{F}\biggl(N_r+N\frac{y}{R'}\biggl)M'+\mbox{(t.d.)}, \label{eq:GRfinal}
\end{align}
where $\mbox{(t.d.)}$ is a total derivative term, we obtain the Hamiltonian density ${\cal H}_{\rm G}$ in the equivalent two-dimensional theory as
\begin{align}
{\cal H}_{\rm G}:=& P_M\dot{M} + P_S\dot{S}-{\cal L} \nonumber \\
   =& N^M M'+N^S P_S,   \label{eq:geometrodynamic hamiltonian}
\end{align}
where we have used Eq.~(\ref{eq:GRfinal}) and defined new Lagrange multipliers $N^M$ and $N^S$ as
\begin{align}
N^M:=&- \frac{\Lambda}{R'}\biggl(N+\frac{y{\dot R}}{F}\biggl) \nonumber\\ 
=&- \frac{\Lambda}{R'}\biggl(N+\frac{y}{F}\left(Ny+N_r R'\right)\biggl) \nonumber\\ 
=&-N\left(\frac{\Lambda}{R'}-\frac{P_M y}{R'}\right) +N_r P_M, 
\label{NM} \\
N^S:=&\dot{S}= \left(Ny+N_r R'\right).\label{NS}
\end{align}

Collecting terms in $N$ and $N_r$, we can express the total Hamiltonian as 
\begin{equation}
{\cal H}_{\rm G}= N\left[\left(\frac{P_M y}{R'} -\frac{\Lambda}{R'}\right)M'+ y P_s\right] + N_r (P_M M' +P_S S').
\label{eq: HG}
\end{equation}
The coefficients of $N$ and $N_r$  are the super-Hamiltonian $H$ and the super-momentum $H_r$, respectively. These can in principle be expressed in terms of Kucha\v{r}' variables by doing the inverse canonical transformation. 
Note that one can replace $H$ by the linear combination of constraints
\begin{equation}
{\cal G} := H - \frac{y}{R'} H_r = -\frac{\Lambda}{R'} M'
\end{equation} 
in agreement with (\ref{M'2}).

The Lagrangian density for the canonical coordinates $(M,S,N^M,N^S)$ is now written as
\begin{align}
{\cal L}= P_M\dot{M} + P_S\dot{S}- N^M M'-N^S P_S, \label{eq:geometrodynamic L}
\end{align}
which corresponds to Eq.~(122) of \cite{kuchar94}. 
The constraints, $M'=0$ and $P_S=0$, are obtained by varying the Lagrange multipliers $N^M$ and $N^S$, respectively. 
On the constraint surface $M=m(t)$ and $P_S=0$ hold.  
The reduced phase space is therefore two-dimensional consisting of $p_m:=\int^\infty_{-\infty} dx P_M(x,t)$ and $m$.
With suitable boundary conditions~\cite{KMT12}, one can repeat the analysis of~\cite{kuchar94} for spacelike slicings that intersect both left and right branches of the outer horizons of eternal black holes to obtain the reduced action:
\begin{align}
I_{(2)} = \int dt \biggl[p_m\dot{m} - ({N}_+-{N}_-)m\biggl],
\label{eq:reduced action}
\end{align}
where $N_{\pm}:=\mp \lim_{x\to \pm\infty}N^M$. The reduced equations of motion in vacuum then imply that $m=m_0=constant$, and $\dot{p}_m = -(N_+-N_-)$.

Lastly, let us derive the boundary term in Eq.~(\ref{boundaryGR}) with the new canonical variables.
Starting from  
\begin{align}
I_{\rm M(GR)}=\int dt\int dx(P_M\dot{M} + P_S\dot{S}- N^M M'-N^S P_S),
\end{align}
we obtain
\begin{align}
\delta I_{\rm M(GR)}=&\int dt\int dx\biggl(\delta P_M\dot{M}+\partial_t(P_M\delta M)-{\dot P}_M\delta {M}+ \delta P_S\dot{S} + \partial_t(P_S\delta  S)-{\dot P}_S\delta  S \nonumber \\
&- \delta N^M M'- (N^M \delta M)'+{N^M}' \delta M-\delta N^S P_S-N^S \delta P_S\biggl) \nonumber \\
=&\int dt\int dx \biggl(\mbox{dynamical terms}\biggl)+\int dx \biggl[P_M\delta M+P_S\delta  S\biggl]_{t=t_1}^{t=t_2}-\int dt\biggl[N^M \delta M\biggl]_{x=-\infty}^{x=+\infty}.
\end{align} 
Under the boundary condition (\ref{bc1})--(\ref{bc6}), we obtain $\delta M \simeq \delta M^\infty(t)$ and $N^M\simeq -N_\infty(t)$ at spacelike infinity and hence we obtain the same result (\ref{boundaryGR}) by setting $\delta M=0$ and $\delta  S=0$ at $t=t_1,t_2$.
One important advantage of the new set of canonical variables is to greatly simplify the calculations.
We will take advantage of this simplification in the next section.

\section{Canonical formalism in Lovelock gravity}
\label{Can form Lovelock}
In this section, we show that all the results in the previous section can be generalized to full Lovelock gravity. 
In particular the transformation from the ADM variables $\{\Lambda, P_\Lambda; R,P_R\}$ to $\{M, P_M; S,P_S\}$ is a well-defined canonical transformation using definitions of $P_M$, $S$, and $P_S$ that are the same as those in general relativity, Eqs.~(\ref{eq:R})--(\ref{eq:Pr1}), and $M$  defined by Eq.~(\ref{qlm-L}).

\subsection{ADM variables}
First we derive the ADM conjugate momenta $P_\Lambda$ and $P_R$.
The Lagrangian density from the action (\ref{action-3})  is 
\begin{align}
{\cal L}=& \frac{(n-2){\cal A}_{n-2}}{2\kappa_n^2}\sum^{[n/2]}_{p=0}{\tilde \alpha}_{(p)}  \Biggl[2pR^{n-2p-1}y(N_r\Lambda)' -\frac{2pN}{n-2p}\biggl((R^{n-2p})'\Lambda^{-1}\biggl)' \nonumber \\
& + (n-2p-1)\biggl\{\left(1-F\right)^{p} +2pF\biggl\}N\Lambda R^{n-2-2p}-2pR^{n-2p-1}y {\dot \Lambda}  \nonumber \\
&+ pR^{n-2p-1}\biggl\{1-(1-F)^{p-1}\biggl\}\biggl\{(\Lambda N_r y+\Lambda^{-1}NR')\frac{F'}{F}-\Lambda y\frac{\dot F}{F} \biggl\}\biggl].\label{Lag-0}
\end{align}
Using the binomial expansion and integration by parts many times, we can rewrite the above Lagrangian density into the following form up to the total derivative. The derivation is presented in Appendix~\ref{appendix2}.
\begin{align}
{\cal L}=& \frac{(n-2){\cal A}_{n-2}}{2\kappa_n^2}\sum^{[n/2]}_{p=0}{\tilde \alpha}_{(p)}  \Biggl[2pR^{n-2p-1}y(N_r\Lambda)'-\frac{2pN}{n-2p}\biggl((R^{n-2p})'\Lambda^{-1}\biggl)' \nonumber \\
& + (n-2p-1)\biggl\{\left(1-F\right)^{p} +2pF\biggl\}N\Lambda R^{n-2-2p}-2pR^{n-2p-1}y {\dot \Lambda}\biggl] \nonumber \\
&-\frac{(n-2){\cal A}_{n-2}}{2\kappa_n^2}\sum^{[n/2]}_{p=2}{\tilde \alpha}_{(p)}\biggl[\sum_{w=0}^{p-2}\frac{p!(-1)^{p-1-w}}{w!(p-1-w)!}\biggl\{ \Lambda N_r yR^{n-2p-1}F^{p-2-w}(\Lambda^{-2}{R'}^2)' \nonumber \\
&+\sum_{j=0}^{p-2-w}\frac{2(p-2-w)!(-1)^{p-2-w-j}}{j!(p-2-w-j)!}\frac{(N_r R^{n-2p-1}\Lambda^{1-2j}{R'}^{2j})'y^{2(p-w-j)-1}}{2(p-w-j)-1} \nonumber \\
&-\frac{(\Lambda^{-1}NR'R^{n-2p-1})'F^{p-1-w}}{p-1-w}\biggl\} +\sum_{w=1}^{p-1}\frac{2p!(-1)^w}{w!(p-1-w)!}R^{n-2p-1} F^{w-1} y\Lambda^{-2}{\dot \Lambda}{R'}^2\nonumber \\
&-\sum_{w=1}^{p-1}\frac{2p!}{w(p-1-w)!}\sum_{j=0}^{w-1}\frac{(-1)^{2w-1-j}}{j!(w-1-j)!} \biggl\{\frac{\partial_t(R^{n-2p-1}\Lambda^{1-2j}){R'}^{2j}y^{2(w-j)+1} }{2(w-j)+1}\nonumber \\
&-j\sum_{q=0}^{2(w-j)+1}\frac{(2w-2j)!(-1)^q}{q!(2w-2j+1-q)!}\frac{{\dot R}^{w-j+1}(R^{n-2p-1}\Lambda^{1-2j}N^{-2(w-j)-1}{R'}^{2j-1+q}N_r^q)'}{2(w-j+1)} \nonumber \\
&-\sum_{q=0}^{2(w-j)-1}\frac{(2w-2j-1)!(-1)^q}{q!(2w-2j-q)!}{\dot R}^{2w-2j-q}(R^{n-2p-1}\Lambda^{-1-2j}N^{-2(w-j)+1}{R'}^{2j+1+q}N_r^q)'\biggl\}\biggl] \label{Lag-1}.
\end{align}

From this Lagrangian density, we obtain 
\begin{align}
P_\Lambda=& -\frac{(n-2){\cal A}_{n-2}}{\kappa_n^2}\biggl[\sum^{[n/2]}_{p=0}{\tilde \alpha}_{(p)}pR^{n-2p-1}y +\sum^{[n/2]}_{p=2}{\tilde \alpha}_{(p)}R^{n-2p-1}y\frac{{R'}^2}{\Lambda^{2}}   \nonumber \\
&\times \sum_{w=1}^{p-1}\frac{p!(-1)^w}{w!(p-1-w)!}\biggl\{F^{w-1}- \sum_{j=0}^{w-1}\frac{(-1)^{w-1-j}}{j!(w-1-j)!} \frac{(1-2j)y^{2(w-j)}}{2(w-j)+1}\frac{{R'}^{2j-2}}{\Lambda^{2j-2}}\biggl\} \biggl]\label{PLambda3}
\end{align}
and
\begin{align}
P_R=& \frac{(n-2){\cal A}_{n-2}}{\kappa_n^2}\sum^{[n/2]}_{p=0}{\tilde \alpha}_{(p)} pR^{n-2p-1} \Biggl[\frac{(N_r\Lambda)'-{\dot \Lambda}}{N}+(n-2p-1)\biggl\{\left(1-F\right)^{p-1} -2\biggl\}\frac{y\Lambda}{R}\biggl] \nonumber \\
&-\frac{(n-2){\cal A}_{n-2}}{2\kappa_n^2}\sum^{[n/2]}_{p=2}{\tilde \alpha}_{(p)} p\biggl[\sum_{w=0}^{p-2}\frac{(p-1)!(-1)^{p-1-w}}{w!(p-1-w)!}N^{-1}\biggl\{2(\Lambda^{-1}NR'R^{n-2p-1})'F^{p-2-w}y\nonumber \\
&+ \sum_{j=0}^{p-2-w}\frac{2(p-2-w)!(-1)^{p-2-w-j}}{j!(p-2-w-j)!}(N_r R^{n-2p-1}\Lambda^{1-2j}{R'}^{2j})'y^{2(p-w-j-1)} \nonumber \\
&+\Lambda N_r R^{n-2p-1}F^{p-3-w}(\Lambda^{-2}{R'}^2)'\biggl(F-2(p-2-w) y^2\biggl) \biggl\} \nonumber \\
&+\sum_{w=1}^{p-1}\frac{2(p-1)!(-1)^w}{w!(p-1-w)!}\biggl\{R^{n-2p-1} \Lambda^{-2}{\dot \Lambda}{R'}^2N^{-1}F^{w-2}\biggl(F-2(w-1) y^2\biggl)\nonumber \\
&- \sum_{j=0}^{w-1}\frac{(-1)^{w-1-j}}{j!(w-1-j)!}\biggl(\partial_t(R^{n-2p-1}\Lambda^{1-2j})\frac{{R'}^{2j}y^{2(w-j)}}{N} +\frac{n-2p-1}{2(w-j)+1}\frac{R^{n-2p-2}{R'}^{2j}y^{2(w-j)+1}}{\Lambda^{2j-1}} \nonumber \\
&-\sum_{q=0}^{2(w-j)+1}\frac{2j(2w-2j)!(-1)^q}{q!(2w-2j+1-q)!}{\dot R}^{2(w-j)+1}(R^{n-2p-1}\Lambda^{1-2j}N^{-2(w-j)-1}{R'}^{2j-1+q}N_r^q)' \nonumber \\
&-\sum_{q=0}^{2(w-j)-1}\frac{(2w-2j-1)!(-1)^q}{q!(2w-2j-1-q)!}{\dot R}^{2w-2j-q-1}(R^{n-2p-1}\Lambda^{-1-2j}N^{-2(w-j)+1}{R'}^{2j+1+q}N_r^q)'\biggl)\biggl\}\biggl].
\end{align}

In general relativity, $P_\Lambda=P_\Lambda[{\dot \Lambda},y({\dot R})]$ and $P_R=P_R[{\dot \Lambda},y({\dot R})]$ can be algebraically solved to give a unique set of ${\dot \Lambda}={\dot \Lambda}[P_\Lambda,P_R]$ and ${\dot R}={\dot R}[P_\Lambda,P_R]$.
In higher-order Lovelock gravity, by contrast, it is not possible to obtain a unique expression in general because of the fact that $P_\Lambda=P_\Lambda[{\dot \Lambda},y({\dot R})]$ and $P_R=P_R[{\dot \Lambda},y({\dot R})]$ are higher-order polynomials of $y$.
As a result, it is difficult to obtain the explicit forms of the super-momentum $H_r$ and the super-Hamiltonian $H$, such that
\begin{align}
{\cal L}= {\dot \Lambda}P_{\Lambda}+{\dot R}P_{R}-N H-N_r H_{r}
\end{align} 
in terms of the ADM variables. However, it is not necessary to do so at this stage. 
Things are greatly simplified by using the generalized Misner-Sharp mass as a new canonical variable.
As we will show, the super-momentum and the super-Hamiltonian with the new set of canonical coordinates are the same as those in general relativity and then the boundary terms at spatial infinity can be easily derived.

\subsection{Generalized Misner-Sharp mass as canonical variable}
We introduce a new set of canonical variables $\{M,P_M;S, P_S\}$ defined in the same way as in general relativity, namely by Eq.~(\ref{qlm-L}), and Eqs.~(\ref{eq:R})--(\ref{eq:Pr1}).
Then, we prove that $\{\Lambda,P_\Lambda;R, P_R\}$ and $\{M,P_M;S, P_S\}$ again satisfy the Liouville form (\ref{Liouville}) with the following total variation and the total derivative terms. The derivation is presented in Appendix~\ref{app:Liouville}.
\begin{align}
\eta:=&\frac{(n-2){\cal A}_{n-2}}{2\kappa_n^2}\biggl[\sum_{p=1}^{[n/2]}{\tilde \alpha}_{(p)}pR^{n-1-2p}\biggl(2y\Lambda-R' \ln\biggl|\frac{R'+y\Lambda}{R'-y\Lambda}\biggl|\biggl) \nonumber \\
&-\sum_{p=2}^{[n/2]}{\tilde \alpha}_{(p)}R^{n-1-2p} \sum_{w=1}^{p-1}\frac{p!}{w(p-1-w)!}\sum_{j=0}^{w-1}\frac{2(-1)^{2w-1-j}{R'}^{2j}y^{2(w-j)+1}}{j!(w-1-j)![2(w-j)+1]\Lambda^{2j-1}}\biggl], \label{deta} \\
\zeta:=&\frac{(n-2){\cal A}_{n-2}}{2\kappa_n^2}\biggl[\sum_{p=1}^{[n/2]}{\tilde \alpha}_{(p)}pR^{n-1-2p}\ln\biggl|\frac{R'+y\Lambda}{R'-y\Lambda}\biggl| +\sum_{p=2}^{[n/2]}{\tilde \alpha}_{(p)}R^{n-1-2p}\nonumber \\
&\times \sum_{w=1}^{p-1}\frac{2p!(-1)^wyR'}{w!(p-1-w)!\Lambda}\biggl\{F^{w-1}+\sum_{j=0}^{w-1}\frac{2j(w-1)!(-1)^{w-1-j}{R'}^{2j-2}y^{2(w-j)}}{j!(w-1-j)![2(w-j)+1]\Lambda^{2j-2}}\biggl\}\biggl]\delta R. \label{zeta'}
\end{align} 

In Appendix~\ref{appendix3}, it is proven that Eq.~(\ref{eq:GRfinal}) still holds in full Lovelock gravity. This immediately implies that the Hamiltonian density in the equivalent two-dimensional theory takes the same form as that in general relativity (\ref{eq:geometrodynamic hamiltonian}), where the definitions of the new Lagrange multipliers $N^M$ and $N^S$ are the same as those in general relativity (\ref{NM}) and (\ref{NS}). 
Finally, the Lagrangian density for the canonical coordinates $\{M,P_M;S,P_S\}$ can be again written as
\begin{align}
{\cal L}= P_M\dot{M} + P_S\dot{S}- N^M M'-N^S P_S \label{eq:geometrodynamic L2}
\end{align}
and the super-Hamiltonian and super-momentum constraints are again as in (\ref{eq: HG}).
 
In comparison to the rather complicated starting point in Eq.~(\ref{L_p}), this equivalent Lagrangian density is extremely simple and the physical meaning of the canonical variables are very clear. 
Remarkably, the coupling constants $\alpha_{(p)}$ do not appear explicitly in (\ref{eq:geometrodynamic L2}).
They are in fact hidden in the definition of the mass function. 
This makes it possible to treat any class of Lovelock gravity in exactly the same way.

\subsection{Fall-off rate at infinity and boundary terms}
In order to prove that the transformation from $\{\Lambda, P_\Lambda,; R,P_R\}$ to $\{M,P_M;S,P_S\}$ is canonical and well-defined, we have to discuss the asymptotic behaviour of the variables.
We adopt the same boundary conditions (\ref{bc1})--(\ref{bc4}) as  in general relativity.
With these conditions, one can verify that the generalized Misner-Sharp mass (\ref{qlm-L}) behaves near spacelike infinity as
\begin{align}
M \simeq \frac{(n-2)A_{n-2}{\tilde \alpha}_{(1)}\Lambda_1(t)}{\kappa_n^2}.
\end{align}  
This is the same as in general relativity and hence we set $\Lambda_1$ as in Eq.~(\ref{bc7}) (where ${\tilde \alpha}_{(1)}=1$) in order that $M\simeq M^\infty(t)$ at infinity.

It can then be shown that the leading terms of $P_\Lambda$, $P_R$, $\zeta$, $\eta$, $P_S$, $P_M$, $N^M$, and $N^S$ are the same as those in the general relativistic case under the boundary condition (\ref{bc1})--(\ref{bc4}).
As a consequence, the proof carries over from general relativity and all the terms in the Liouville form (\ref{Liouville}) are well behaved  at spacelike infinity.
This is sufficient to prove the transformation from $\{\Lambda, P_\Lambda,; R,P_R\}$ to $\{M,P_M;S,P_S\}$ is indeed a well-defined canonical transformation.
and that the Hamiltonian $\int^\infty_{-\infty}dx(N^M M'+N^S P_S)$ is also finite.

We now have the following two-dimensional action with a new set of canonical variables;
\begin{align}
I_{\rm M(L)}=\int dt\int dx(P_M\dot{M} + P_S\dot{S}- N^M M'-N^S P_S),
\end{align}
with the same asymptotic behavior as in general relativity. The boundary term for the above action that makes the variational principle well defined is then also the same as in general relativity:
\begin{align}
\delta I_{\rm M(L)}=\int dt\int dx \biggl(\mbox{dynamical terms}\biggl)+\int dt\biggl[N_\infty(t)\delta M^\infty(t)\biggl]_{x=-\infty}^{x=+\infty}.
\end{align} 

Given the above, we can now write down the super-Hamiltonian and super-momentum constraints for full Lovelock gravity. In terms of the geometrodynamical variables they are the same expressions as in general relativity:
\bea
H &=& \left(\frac{P_M y}{R'} -\frac{\Lambda}{R'}\right)M'
+ y P_s\, ,
\label{eq:super H} \\
H_r &=& P_M M' +P_S S'.
\label{eq:super Hr}
\eea
The expressions in terms of ADM variables are considerably more complicated and can in principle be obtained once again by substution from Eqs.~(\ref{eq:R})--(\ref{eq:Pr1}), with $M$  defined by Eq.~(\ref{qlm-L}).

\section{Adding matter fields}
\label{matter}
In this section, we introduce matter fields in the argument with the ADM variables discussed in the previous sections.
Here we write super-momentum and super-Hamiltonian for gravity as $H_r^{\rm (G)}$ and $H^{\rm (G)}$ in order to distinguish from the total super-momentum and super-Hamiltonian including matter contributions.
The following argument is valid in full Lovelock gravity.

It can be shown from Eqs.~(\ref{defy}), (\ref{eq:PM}), (\ref{eq:R}), (\ref{eq:Pr1}) and (\ref{eq:GRfinal}) that the gravitational Hamiltonian $H_{\rm G}$ is given by:
\be
\label{HGADM}
H_{\rm G} = \int dx (NH^{\rm (G)} + N_rH_r^{\rm (G)}),
\ee
where
\begin{align}
\label{momconst1}
H_r^{\rm (G)} =& P_S S^\prime + P_M M^\prime = P_R R^\prime  - P_\Lambda^\prime \Lambda,\\
\label{Hconst1}
H^{\rm (G)} =& -\frac{\Lambda}{R^\prime}M^\prime  + \frac{y}{R^\prime}H_r^{(G)}.
\end{align}
We have used (\ref{M'2}) to derive (\ref{Hconst1}).
Since Eq.~(\ref{PLambda3}) shows that $y$ is not a function of $N$ or $N_r$, we can see that the Hamiltonian density is the sum of Lagrange multiplier times constraints.

\subsection{Massless scalar field}
\label{Scalar field}
First we consider a massless scalar field $\psi$ as a matter field, of which action is $I_{\rm matter}=I_\psi$ in the action (\ref{action}):
\begin{align}
I_\psi = -\frac{1}{2}\int d^{n}x \sqrt{-g} (\nabla \psi)^2.
\end{align}
The equivalent two-dimensional action in the symmetric spacetime under consideration is given by 
\begin{align}
\label{scalaraction2}
I_\psi &= -\frac{{\cal A}_{n-2}}{2}\int d^2{\bar y} \sqrt{-\g} R^{n-2} (D \psi)^2 \\ \nonumber
& = -\frac{{\cal A}_{n-2}}{2}\int dxdt \frac{\Lambda R^{n-2} }{N} \left( -\psid^2 + 2N_r\psip \psid + (N^2\Lambda^{-2}-N_r^2) \psip^2 \right).
\end{align}
This gives the momentum conjugate $\Pis$ to $\psi$ as
\be
\label{Pip}
\Pis =  \frac{{\cal A}_{n-2}\Lambda R^{n-2}}{N} \left( \psid - N_r \psip \right),
\ee
with which we can write the matter action as
\begin{align}
\label{scalaraction3}
I_\psi=& \int dxdt \psid \Pis- \int dxdt N \left[ \frac{1}{2\Lambda} \left( \frac{\Pis^2}{{\cal A}_{n-2} R^{n-2}} +{\cal A}_{n-2}R^{n-2} \psip^2 \right) + \Pis \psip\frac{N_r}{N}  \right].
\end{align}

Equation (\ref{scalaraction2}) does not contain any derivatives of the metric or $R$, which means that adding the scalar action to the gravitational action (\ref{eq:geometrodynamic hamiltonian}) does not change $P_\Lambda$ or $P_R$.  
This allows us to write the total Hamiltonian as the sum of the gravitational and matter parts.  
Using Eqs.~(\ref{HGADM}), (\ref{momconst1}), (\ref{Hconst1}), and (\ref{scalaraction3}), we obtain the total Hamiltonian $H_{\rm total} $ as
\begin{align}
\label{H8}
H_{\rm total} & = \int dx N\Bigg[-\frac{\Lambda}{R^\prime}M^\prime  + \frac{y}{R^\prime}H_r  + \frac{N_r}{N} H_r + \frac{1}{2\Lambda} \left( \frac{P_\psi^2}{{\cal A}_{n-2} R^{n-2}} + {\cal A}_{n-2} R^{n-2} \psip^2 \right) +P_\psi \psip\frac{N_r}{N}   \Bigg] \none
& = \int dx \biggl[ N(H^{\rm (G)}+H^{\rm (M)}) + N_r(H_r^{\rm (G)}+H_r^{\rm (M)}) \biggl],
\end{align}
where $y$ is a function of the phase space variables, $\Lambda$, $P_\Lambda$ and $R$ via Eq.~(\ref{PLambda3}).  
The super-Hamiltonian $H^{\rm (M)}$ and super-momentum $H_r^{\rm (M)}$ for $\psi$ are given by
\begin{align}
\label{HconstM}
H^{\rm (M)} =& \frac{1}{2\Lambda} \left( \frac{P_\psi^2}{{\cal A}_{n-2} R^{n-2}} +{\cal A}_{n-2}R^{n-2} \psip^2 \right),\\
\label{momconstM}
H_r^{\rm (M)} =& P_\psi \psip.
\end{align}
The Poisson bracket of 
 Hamiltonian constraint, $H = H^{\rm (G)} + H^{\rm (M)}$ with the total momentum constraint, $H_r = H_r^{\rm (G)} + H_r^{\rm (M)}$ is given by
\be
\label{firstclasstotal}
\{H, H_r \} = \{H^{\rm (G)}, H_r^{\rm (G)} \} + \{H^{\rm (G)}, H_r^{\rm (M)} \} + \{H^{\rm (M)}, H_r^{\rm (G)} \} + \{H^{\rm (M)}, H_r^{\rm (M)} \}.
\ee
Because our theory is diffeomorphism invariant, $\{H, H_r \}$ must be weekly equal to zero.  

Because there are two first class constraints, there are two gauge choices to pick.  
We choose our first gauge as
\be
\label{gauge1}
\chi := R - x \approx 0.
\ee
This forces the spatial coordinate to be the areal radius which means that $R$ is no longer a phase space variable, it is now a coordinate.  
In order to insist that $\chi$ is satisfied at every time slice, we must insist that $\dot{\chi}=\{\chi,H\} \approx 0$, which shows $N_r/N +y/R^\prime \approx 0$.  
We use this relation to write one Lagrange multiplier in terms of the other.  
This leaves us with one Lagrange multiplier which reflects the fact that there is only one gauge fix left to choose.

We can now plug the gauge choice (\ref{gauge1}) and its consistency condition into the Hamiltonian as long as we use Dirac brackets to evaluate the equations of motion in the end.  
Note that the remaining phase space variables, $\Lambda$, $P_\Lambda$, $\psi$ and $P_\psi$, all commute with $\chi$ and so the Poisson bracket is the same as the Dirac bracket.  
Plugging $\chi = \dot{\chi} = 0$ into Eq.~(\ref{H8}) gives
\be
\label{H8b}
H_{\rm total} = \int dR N \Biggl[-\Lambda M^\prime + \frac{1}{2\Lambda} \left( \frac{\Pis^2}{{\cal A}_{n-2}R^{n-2}} + {\cal A}_{n-2}R^{n-2} \psip^2 \right) - y \Pis \psip \Biggr].
\ee
In the last term we replaced $N_r/N$ by $-y$ as required.  
Since the mass equation (\ref{qlm-L}) is written as
\begin{align}
\label{Massfunc}
M=\frac{(n-2){\cal A}_{n-2}}{2\kappa_n^2}\sum_{p=0}^{[n/2]} \tilde{\alpha}_{(p)}R^{n-1-2p}\left(1-\Lambda^{-2} + y^2\right)^p,
\end{align}
we can write $y$($=N_r/N$) in terms of the mass function.  
For this reason we leave the factor of $N_r/N$ in the Hamiltonian with the understanding that it is the solution to Eq.~(\ref{Massfunc}).

For our second gauge choice we choose
\be
\label{gauge2}
\xi := \Lambda - 1\approx 0.
\ee
By the same reasoning used for the first gauge choice we can set $\xi$ strongly to zero (namely since $\Lambda$ commutes with $\psi$ and $P_\psi$) which gives the Hamiltonian
\be
\label{H8d}
H_{\rm total} = \int dR N \Biggl[ -M^\prime + \frac{1}{2} \left( \frac{\Pis^2}{{\cal A}_{n-2} R^{n-2}} + {\cal A}_{n-2} R^{n-2} \psip^2 \right) + \Pis \psip  \frac{N_r}{N}  \Biggr]
\ee
and the mass function
\begin{align}
\label{Nsig}
M= \frac{(n-2){\cal A}_{n-2}}{2\kappa_n^2}\sum_{p=0}^{[n/2]} \tilde{\alpha}_{(p)}R^{n-1-2p}\left(\frac{N_r}{N}\right)^{2p}.
\end{align}
To see the significance of this gauge choice, notice from Eq.~(\ref{qlm-L}) that $g^{11} \to 1-2\kappa_n^2M/[(n-2){\ma A_{(n-2)}}\tilde{\alpha}_{(1)}R^{n-3}]$ in the general relativistic case when we strongly set $\xi$ and $\chi$ to zero.  This gives the metric in the non-static version of Painlev\'{e}-Gullstrand coordinates:
\be
ds_{(2)}^2= -N^2 \left( 1-\frac{2\kappa_n^2M}{(n-2){\ma A_{(n-2)}}\tilde{\alpha}_{(1)}R^{n-3}} \right)  dt^2 + 2 N \sqrt{\frac{2\kappa_n^2M}{(n-2){\ma A_{(n-2)}}\tilde{\alpha}_{(1)}R^{n-3}}} dtdR + dR^2.
\ee

To ensure that the second gauge condition is conserved in time we must insist that $d(\Lambda - 1)/dt = \{\Lambda - 1, H\} = 0 \rightarrow \delta H/\delta P_\Lambda = 0$.  
Although we have chosen to write $N_r/N$ in terms of the mass function, it can also be written in terms of $P_\Lambda$.  
All of the $P_\Lambda$ dependence in the Hamiltonian is in the terms of $N_r/N$.  
Therefore we can write
\begin{align}
\frac{\delta H_{\rm total}}{\delta P_\Lambda} & = \frac{\delta}{\delta P_\Lambda} \int dR \left( N^\prime M + N P_\psi \psip  \frac{N_r}{N}  \right) \nonumber \\ 
& = N^\prime \frac{\partial M}{\partial (N_r/N)} \frac{\partial (N_r/N)}{\partial P_\Lambda} + N P_\psi \psip\frac{\partial (N_r/N)}{\partial P_\Lambda},  \label{consistency conditon 2}
\end{align}
from which the consistency condition is given as
\be
\label{sigma1}
N^\prime \frac{\partial M}{\partial (N_r/N)} + N P_\psi \psip = 0,
\ee
where it is understood that we use (\ref{Nsig}) to find $\partial M/\partial (N_r/N)$ and write $N_r/N$ in terms of the mass function $M$.  
Notice that the actual relation between $N_r/N$ and $P_\Lambda$ is not needed.

Using Hamilton's equations and Eq.~(\ref{H8d}), we find
\begin{align}
\label{psidot}
\dot{\psi} =& N \left( \frac{\Pis}{{\cal A}_{n-2} R^{n-2}} + \psip\frac{N_r}{N}  \right),\\
\label{Pisdot}
\dot{P}_\psi =& \left[ N \left( {\cal A}_{n-2} R^{n-2} \psip +  \Pis \frac{N_r}{N}  \right) \right]^\prime.
\end{align}
These equations, along with the consistency conditions (\ref{Nsig}) and (\ref{sigma1}) and the Hamiltonian constraint
\be
\label{Cconstraint}
-M^\prime + \frac{1}{2} \left( \frac{\Pis^2}{{\cal A}_{n-2}R^{n-2}} + {\cal A}_{n-2} R^{n-2} \psip^2 \right) + \Pis \psip  \frac{N_r}{N}= 0,
\ee
determine the evolution of a collapsing scalar field.

\subsection{Charged scalar field}
In this subsection, we consider a U(1) gauge field $A_\mu$ coupled to a charged complex massless scalar field $\psi = (\psi_1 + i\psi_2)/\sqrt{2}$, where $\psi_1$ and $\psi_2$ are real functions.
We write the action for this matter as $I_{\rm matter}=I_{\rm EM}$:
\be
\label{EMaction1}
I_{\rm EM}= \int d^nx \sqrt{-g}\left[-\left( \partial^\mu + ieA^\mu \right) \psi^* \left( \partial_\mu - ieA_\mu \right)  \psi- \frac{1}{4} F^{\mu \nu} F_{\mu \nu} \right],
\ee
where $e$ is the charge and the Faraday tensor $F_{\mu \nu}$ is defined in terms of the gauge field as $F_{\mu \nu}=\partial_\mu A_\nu-\partial_\nu A_\mu$.  
Under the symmetry assumption in the present paper both for gravity and matter, the equivalent two-dimensional action is given by 
\be
\label{EMaction2}
I_{\rm EM}={\cal A}_{n-2}\int dtdx \sqrt{-g_{(2)}} R^{n-2}\left[-\left( \partial^B + ieA^B \right) \psi^* \left( \partial_B - ieA_B \right)  \psi -\frac{1}{4} F^{AB} F_{AB}\right].
\ee
Adopting the ADM coordinates, we obtain
\begin{align}
\label{EMaction3}
I_{\rm EM} =&\frac{{\cal A}_{n-2}}{2} \int dtdx \frac{R^{n-2}\Lambda}{N}\Bigg[(\dot{\psi}_1^2 + \dot{\psi}_2^2) - 2N_r(\dot{\psi_1}\psi_1^\prime + \dot{\psi_2}\psi_2^\prime) +(N_r^2-N^2\Lambda^{-2})(\psi_1^{\prime 2} + \psi_2^{\prime 2}) \nonumber \\
& -2e \biggl\{ (A_0-N_rA_1)(\dot{\psi}_2 \psi_1 - \dot{\psi}_1 \psi_2) - \biggl(N_r(A_0-N_rA_1) + N^2\Lambda^{-2}A_1\biggl)(\psi_2^\prime \psi_1 - \psi_1^\prime \psi_2) \biggl\}\nonumber \\ 
& +e^2 \biggl((A_0-N_rA_1)^2-N^2\Lambda^{-2}A_1^2\biggl)(\psi_1^2 + \psi_2^2) + \Lambda^{-2} (\dot{A_1} - A_0^\prime)^2 \Bigg],
\end{align}
where $A_\mu dx^\mu=A_0(t,x)dt+A_1(t,x)dx$.
From the above action we find the conjugate momenta:
\begin{align}
\label{Ppsi1}
P_{\psi 1} =& \frac{{\cal A}_{n-2}R^{n-2}\Lambda}{N}\left[ \dot{\psi}_1 - N_r \psi_1^\prime +e(A_0 - N_r A_1) \psi_2 \right],\\
\label{Ppsi2}
P_{\psi 2} =&\frac{{\cal A}_{n-2}R^{n-2}\Lambda}{N}\left[ \dot{\psi}_2 - N_r \psi_2^\prime -e(A_0 - N_r A_1) \psi_1 \right],\\
\label{PA0}
P_{A0} =& 0,\\
\label{PA1}
P_{A1} =& \frac{{\cal A}_{n-2}R^{n-2}(\dot{A_1} - A_0^\prime)}{N\Lambda},
\end{align}
which give the Hamiltonian for the present matter field:
\begin{align}
\label{HEM1}
H_{\rm EM}=&\int dx \Bigg[ \frac{N}{2{\cal A}_{n-2}\Lambda R^{n-2}}(P_{\psi 1}^2 + P_{\psi 2}^2) + e(A_0-N_rA_1)(P_{\psi 2} \psi_1 - P_{\psi 1} \psi_2) \nonumber \\
& + N_r(P_{\psi 1} \psi_1^\prime + P_{\psi 2} \psi_2^\prime) + \frac{NR^{n-2}}{2{\cal A}_{n-2}\Lambda}\biggl\{ (eA_1\psi_1-\psi_2^\prime)^2+(eA_1\psi_2+\psi_1^\prime)^2 \biggl\} \nonumber \\
& + \frac{N\Lambda}{2{\cal A}_{n-2}R^{n-2}}P_{A1}^2 + P_{A1}A_0^\prime \Bigg].
\end{align}
Since the action (\ref{EMaction3}) contains no derivatives of the metric or $R$, the addition of $I_{\rm EM}$ to the gravitational action does not alter the Hamiltonian analysis and allows us to write the total Hamiltonian as
\be
\label{HEMtotal1}
H_{\rm total} = \int dx \biggl[N(H^{\rm (G)} + H^{\rm (EM)}) + N_r(H_r^{\rm (G)} + H_r^{\rm (EM)}) + A_0H_{A0}^{\rm (EM)} \biggl],
\ee
where $H^{\rm (G)}$, $H_r^{\rm (G)}$ and $H_{A0}^{\rm (EM)}$ are given by Eqs.~(\ref{Hconst1}) and (\ref{momconst1}) and $H^{\rm (EM)}$ and $H_r^{\rm (EM)}$ are given by
\begin{align}
H^{\rm (EM)}=& \frac{P_{\psi 1}^2 + P_{\psi 2}^2}{2{\cal A}_{n-2}\Lambda R^{n-2}} +\frac{{\cal A}_{n-2}R^{n-2}}{2\Lambda}\biggl[ (eA_1\psi_1-\psi_2^\prime)^2+(eA_1\psi_2+\psi_1^\prime)^2 \biggl]+ \frac{\Lambda P_{A1}^2}{2 {\cal A}_{n-2}R^{n-2}},\label{HEM}\\
\label{HrEM}
H_r^{\rm (EM)}=& -eA_1(P_{\psi 2} \psi_1 - P_{\psi 1} \psi_2) + (P_{\psi 1} \psi_1^\prime + P_{\psi 2} \psi_2^\prime),\\
\label{HA0EM}
H_{A0}^{\rm (EM)}=&e(P_{\psi 2} \psi_1 - P_{\psi 1} \psi_2)-P_{A1}^\prime,
\end{align}
where we used integration by parts and asymptotic condition $P_
{A1}A_0\to 0$ at infinity to derive Eq.~(\ref{HA0EM}).
The consistency condition on the constraint (\ref{PA0}) is $\{P_{A0},H_{\rm total}\} = 0$, which gives $e(P_{\psi 2} \psi_1 - P_{\psi 1} \psi_2)-P_{A1}^\prime=0$.  This condition is already added into the Hamiltonian with $A_0$ as its Lagrange multiplier.  
Since $P_{A0}$ is weekly equal to zero we can use the equation of motion for $P_{A0}$ to show that $H_{A0}^{\rm (EM)}$ is weekly equal to zero and is, therefore, a constraint in the same way as the constraints multiplying $N$ and $N_r$.

This Hamiltonian is composed of three first class constraints which means that there are three gauge choices to make.  
Our first two gauge choices will be the same as in section~\ref{Scalar field}.  
Using similar reasoning we can write the Hamiltonian as
\begin{align}
H_{\rm total} =& \int dR \biggl[ N\biggl\{ -M^\prime +\frac{P_{\psi 1}^2 + P_{\psi 2}^2}{2{\cal A}_{n-2}R^{n-2}} +\frac{{\cal A}_{n-2}R^{n-2}}{2}\biggl( (eA_1\psi_1-\psi_2^\prime)^2 + (eA_1\psi_2+\psi_1^\prime)^2 \biggl)  \nonumber \\ 
& +\frac{P_{A1}^2}{2{\cal A}_{n-2}R^{n-2}} + \frac{N_r}{N} \biggl( -eA_1(P_{\psi 2} \psi_1 - P_{\psi 1} \psi_2) + (P_{\psi 1} \psi_1^\prime + P_{\psi 2} \psi_2^\prime) \biggl) \biggl\} \nonumber \\ 
& + A_0\biggl(e(P_{\psi 2} \psi_1 - P_{\psi 1} \psi_2)-P_{A1}^\prime\biggl)\biggl]. \label{HEMtotal1b}
\end{align}
Just as in section \ref{Scalar field} the consistency condition on the first gauge fix requires us to write $N_r/N$ as a function of $M$ using Eq.~(\ref{Nsig}). 
The consistency condition on the second gauge choice, analogous to Eq.~(\ref{sigma1}), is given by
\be
\label{EM consistency condition 2}
N^\prime \frac{\partial M}{\partial (N_r/N)} + N \biggl( -eA_1(P_{\psi 2} \psi_1 - P_{\psi 1} \psi_2) + (P_{\psi 1} \psi_1^\prime + P_{\psi 2} \psi_2^\prime) \biggl) = 0.
\ee
For our third gauge we choose
\be
\label{A0gauge}
\epsilon:=A_1 \approx 0,
\ee
which is the coulomb gauge with the constant, $A_1$ chosen to be zero.  This condition, along with the electromagnetic constraint,
\be
\label{EMConstraint}
e(P_{\psi 2} \psi_1 - P_{\psi 1} \psi_2)-P_{A1}^\prime \approx 0,
\ee
removes $A_1$ and its conjugate momentum $P_{A1}$ from the set of phase space variables.  
We can therefore set $\epsilon$ strongly to zero in the Hamiltonian as we did for the first two gauge choices.  
This gives the following Hamiltonian:
\begin{align}
H_{\rm total} =& \int dR \biggl[ N\biggl\{ -M^\prime +\frac{P_{\psi 1}^2 + P_{\psi 2}^2}{2{\cal A}_{n-2}R^{n-2}} +\frac{{\cal A}_{n-2}R^{n-2}}{2}(\psi_2^{\prime2} + \psi_1^{\prime2} )  \nonumber \\ 
& +\frac{P_{A1}^2}{2{\cal A}_{n-2}R^{n-2}} + \frac{N_r}{N} (P_{\psi 1} \psi_1^\prime + P_{\psi 2} \psi_2^\prime) \biggl\} + A_0\biggl(e(P_{\psi 2} \psi_1 - P_{\psi 1} \psi_2)-P_{A1}^\prime\biggl)\biggl], \label{HEMtotal2}
\end{align}
where it is understood that $P_{A1}$ is the solution of Eq.~(\ref{EMConstraint}).  
The consistency condition on Eq.~(\ref{A0gauge}) is given by
\begin{align}
\{\epsilon , H_{\rm total} \} \approx 0 &\to A_0^\prime + \frac{NP_{A1}}{{\cal A}_{n-2}R^{n-2}} \approx 0 \nonumber \\
&\to A_0^\prime \approx  -\frac{eN}{{\cal A}_{n-2}R^{n-2}} \int dR (P_{\psi 2} \psi_1 - P_{\psi 1} \psi_2),\label{EM consistency condition 3}
\end{align}
which puts a condition on the final Lagrange multiplier and must be satisfied at every time slice.  
This is the last consistency condition on $\epsilon$.

With the fully gauge fixed Hamiltonian (\ref{HEMtotal2}) we may write down Hamilton's equations of motion in terms of the remaining phase space variables, $\psi_1$, $P_{\psi1}$, $\psi_2$ and $P_{\psi2}$.  
The equations of motion are given by
\begin{align}
\label{EMpsi1EOM}
\dot{\psi_1} =& N\left(\frac{P_{\psi1}}{{\cal A}_{n-2}R^{n-2}} + \frac{N_r}{N} \psi_1^\prime\right) - eA_0\psi_2,\\
\label{EMpsi2EOM}
\dot{\psi_2} =& N\left(\frac{P_{\psi2}}{{\cal A}_{n-2}R^{n-2}} + \frac{N_r}{N} \psi_2^\prime\right) + eA_0\psi_1,\\
\label{EMP1EOM}
\dot{P}_{\psi1} =& \left[N \left({\cal A}_{n-2}R^{n-2} \psi_1^\prime + \frac{N_r}{N} P_{\psi1} \right) \right]^\prime - eA_0 P_{\psi2},\\
\label{EMP2EOM}
\dot{P}_{\psi2} =& \left[N \left({\cal A}_{n-2}R^{n-2} \psi_2^\prime +\frac{N_r}{N} P_{\psi2} \right) \right]^\prime + e A_0 P_{\psi1}.
\end{align}
It must be remembered that at every time slice the equations of motion must be supplemented by the consistency conditions (\ref{Nsig}), (\ref{EM consistency condition 2}), and (\ref{EM consistency condition 3}), as well as the Hamiltonian constraint:
\be
\label{EMNconstraint}
-M^\prime + \frac{P_{\psi 1}^2 + P_{\psi 2}^2}{2{\cal A}_{n-2}R^{n-2}} + \frac{{\cal A}_{n-2}R^{n-2}}{2}(\psi_2^{\prime2} + \psi_1^{\prime2}) + \frac{P_{A1}^2}{2{\cal A}_{n-2}R^{n-2}} + \frac{N_r}{N} (P_{\psi 1} \psi_1^\prime + P_{\psi 2} \psi_2^\prime) = 0,
\ee
where it is understood that $P_{A1}$ is the solution of Eq.~(\ref{EMConstraint}).

\section{Conclusions}
\label{Conclusions}
In this paper, we have performed the Hamiltonian analysis for spherically symmetric spacetimes in general Lovelock gravity in arbitrary dimensions.
We have shown that, as in general relativity, the areal radius and the generalized Misner-Sharp quasi-local mass $M$ are natural canonical variables that yield the remarkably simple, geometrical action (\ref{eq:geometrodynamic L2}) for the generic theory. Using these variables also enabled us to rigorously derive the super-Hamiltonian and super-momentum constraints (\ref{eq:super H}) and (\ref{eq:super Hr}) for the most general theory, a task that would have been daunting at best, if not impossible, in terms of ADM variables. Most importantly, our results are useful: the geometrodynamic variables allow the physical phase space of the vacuum theory to be explicitly parametrized in terms of the ADM mass and its conjugate momentum, as done for general relativity by Kucha\v{r}~\cite{kuchar94}. This in turn provides a rigorous starting point for the quantization of  Lovelock black holes using the techniques of \cite{Louko1996}. Finally, the simple form of the Hamiltonian allows us to gauge fix and derive the Hamiltonian equations of motion for the collapse of self gravitating matter in flat slice coordinates. We are now in a position to study the dynamics of black hole formation in generic Lovelock gravity. This is currently in progress. 

Finally we note that while the equations of motion remain virtually unchanged for the case of non-compact symmetry group (i.e. $k=0,-1$) we have nonetheless focused on spherical symmetry ($k=1$). This is perhaps the most interesting case because it describes physically relevant asymptotically flat black holes. The non-compact cases are also of interest in part because of the connection with the AdS/CFT correspondence, for example. However the boundary conditions for $k=0,-1$ require more detailed analysis. Moreover, it has only been rigorously proven that dimensional reduction commutes with the variational principle for spherically symmetric space-times. For these reasons we defer consideration of $k=0,-1$ to a separate work.

{\bf Acknowledgments}
HM thanks Jorge Zanelli for discussions. 
This research was supported in part by the Natural Sciences and Engineering Research Council of Canada and by the JSPS Grant-in-Aid for Scientific Research (A) (22244030). This work has been partially funded by the Fondecyt grants 1100328, 1100755 and by the Conicyt grant "Southern Theoretical Physics Laboratory" ACT-91. 
The Centro de Estudios Cient\'{\i}ficos (CECs) is funded by the Chilean Government through the Centers of Excellence Base Financing Program of Conicyt. TT thanks the University of Manitoba for funding. GK and TT are very grateful to CECs, Chile for its kind hospitality during the initial stages of this work. HM would like to thank University of Winnipeg, for its hospitality while part of this work was carried out. The authors thank Julio Oliva for bringing Ref.\cite{Palais1979} to their attention. 
GK is grateful to Jorma Louko for introducing him to the beauty of Kucha\v{r}' geometrodynamics. TT and GK also thank Danielle Leonard and Robert Mann for the collaboration in~\cite{TLKM} which was instrumental in leading to the present work.

\appendix

\section{Derivations}
\label{app:Derivations}
In this appendix, we present the details of several lengthy derivations of results in the main text.
The following relations will be useful for much of the following:
\begin{align}
y\Lambda \delta F=&y\Lambda \biggl(\frac{\delta({R'}^2)}{\Lambda^2}-2y\delta y\biggl)-2y(y^2+F)\delta \Lambda, \label{useful0}\\
y\Lambda \frac{\delta F}{F}=&2\delta(y\Lambda)-2y\delta \Lambda-R'\delta \biggl(\ln\biggl|\frac{R'+y\Lambda}{R'-y\Lambda}\biggl|\biggl). \label{useful1}
\end{align}
Equation (\ref{useful1}) can be written as
\begin{align}
y\Lambda \frac{\delta F}{F}=&y\Lambda \frac{\delta(({R'}^2-y^2\Lambda^2)/\Lambda^2)}{({R'}^2-y^2\Lambda^2)/\Lambda^2}  \nonumber \\
=&-2(y\Lambda)^2\frac{\delta(y\Lambda)}{{R'}^2-(y\Lambda)^2}-2y\delta \Lambda+y\Lambda \frac{2R'\delta R'}{{R'}^2-y^2\Lambda^2} \nonumber \\
=&2\delta(y\Lambda)-2y\delta \Lambda-2{R'}^2\frac{\delta(y\Lambda)}{{R'}^2-(y\Lambda)^2}+y\Lambda \frac{2R'\delta R'}{{R'}^2-y^2\Lambda^2} \nonumber \\
=&2\delta(y\Lambda)-2y\delta \Lambda-R'\delta \biggl(\ln\biggl|\frac{R'+y\Lambda}{R'-y\Lambda}\biggl|\biggl).
\end{align}
Equations (\ref{useful0})  and (\ref{useful1}) will also be used  by replacing $\delta$ by $\partial_t$ or $\partial_x$.

\subsection{Lagrangian density~(\ref{eq:lovelock simplified})}
\label{appendix1}
We now derive the Lagrangian density~(\ref{eq:lovelock simplified}) from Eq.~(\ref{L_p}).
Using the binomial expansion for the last two terms in (\ref{L_p}) yields
\begin{align}
& 2p(p-1) \frac{(D^2R)^2 - (D^AD_BR)(D^BD_AR)}{R^2} \left(\frac{k-(DR)^2}{R^2}\right)^{p-2} + p\overset{(2)}{\cal R} \left(\frac{k-(DR)^2}{R^2}\right)^{p-1} \nonumber \\
=&R^{-2(p-1)} \biggl[pk^{p-1}\overset{(2)}{\cal R}+\{(D^2R)^2 - (D^AD_BR)(D^BD_AR)\}\sum_{i=0}^{p-2}\frac{2(i+1)p!k^{p-2-i}(-1)^i(DR)^{2i} }{(i+1)!(p-2-i)!}\nonumber \\
& - (D^A R)\left(D^2D_AR - D_AD^2R\right)\sum_{i=0}^{p-2}\frac{2p!k^{p-2-i}(-1)^{i}(DR)^{2i} }{(i+1)!(p-2-i)!}\biggl],
\end{align}
where we used the following two-dimensional identity:
\begin{align}
(D R)^2 \overset{(2)}{\cal R}\equiv  2(D^A R)\left(D^2D_AR - D_AD^2R\right). 
\end{align}
This identity can be derived from Eq.~(2.10) of \cite{TLKM}.

Eq.~(\ref{L_p}) now reduces to
\begin{align}
\label{eq:lp1}
{\cal L}_{(p)} =& \frac{(n-2)!}{(n-2p)!} \biggl[(n-2p)(n-2p-1)\left(\frac{k-(DR)^2}{R^2}\right)^{p} -2p(n-2p)\frac{D^2R}{R}\left(\frac{k-(DR)^2}{R^2}\right)^{p-1} \nonumber \\
& + pk^{p-1}R^{2-2p}\overset{(2)}{\cal R}  + \sum_{i=0}^{p-2} \frac{2(-1)^ik^{p-2-i}p! (DR)^{2i}}{(i+1)!(p-2-i)!} \biggl\{(i+1)\frac{(D^2R)^2 - (D^AD_BR)(D^BD_AR)}{R^{2p-2}} \nonumber \\
&- \frac{D^AR(D^2D_AR - D_AD^2R)}{R^{2p-2}} \biggl\}\biggl].
\end{align}
Using integration by parts, we can rewrite the term in curly brackets in (\ref{eq:lp1}) as
\begin{align}
&\sum_{i=0}^{p-2} \frac{2(-1)^ik^{p-2-i}p! (DR)^{2i}}{(i+1)!(p-2-i)!} R^{n-2p}\biggl\{(i+1)[(D^2R)^2 - (D^AD_BR)(D^BD_AR)] \nonumber \\
&- D^AR(D^2D_AR - D_AD^2R)\biggl\} \nonumber \\
=&\sum_{i=0}^{p-2} \frac{2(-1)^ik^{p-2-i}p! }{(i+1)!(p-2-i)!}(DR)^{2i}D_A(R^{n-2p}) \biggl\{ \frac12D^A((DR)^2)- (D^2R)(D^AR)\biggl\}+\mbox{(t.d.)},
\end{align}
where we used the following identity:
\begin{align}
(D^2R)^2 - (D^AD_BR)(D^BD_AR) + \frac{(D(DR)^2)^2}{2(DR)^2} - \frac{(D^AR)(D_A(DR)^2)(D^2R)}{(DR)^2} \equiv 0.
\end{align}
Using the above result together with the integration by parts and the following identity;
\begin{align}
\sum_{i=0}^{p-2} \frac{2(-1)^ik^{p-2-i}p! (DR)^{2i}}{(i+1)!(p-2-i)!}=&-\sum_{w=1}^{p-1} \frac{2(-1)^{w}k^{p-1-w}p! (DR)^{2w}(DR)^{-2}}{w!(p-1-w)!} \nonumber \\
=&-\sum_{w=0}^{p-1} \frac{2(-1)^{w}k^{p-1-w}p! (DR)^{2w}(DR)^{-2}}{w!(p-1-w)!}+2pk^{p-1} (DR)^{-2} \nonumber \\
=&\frac{2pk^{p-1}-2p(k-(DR)^2)^{p-1}}{(DR)^{2}},
\end{align}
we can rewrite expression (\ref{eq:lp1}) in the form (\ref{eq:lovelock simplified}) up to a total derivative.

\subsection{Lagrangian density (\ref{Lag-1})}
\label{appendix2}
In this appendix, we show how to derive the Lagrangian density (\ref{Lag-1}) from the action (\ref{action-3}).
{ While we consider only the spherically symmetric case $(k=1$) in the main text, here we derive the equations for general $k$.}

For this purpose, we separate the action (\ref{action-3})  into two portions:
\begin{align}
I_{\rm M}=&I_1+I_2+\mbox{(t.d.)}, \label{I_M} \\
I_1:=& \frac{(n-2)V_{n-2}^{(k)}}{2\kappa_n^2}\sum^{[n/2]}_{p=0}\int d^2{\bar y}{\tilde \alpha}_{(p)}\Biggl[2pk^{p-1}R^{n-2p-1}y(N_r\Lambda)'-\frac{2pk^{p-1}N}{n-2p}\biggl((R^{n-2p})'\Lambda^{-1}\biggl)'  \nonumber \\
&+ pR^{n-2p-1}\biggl\{k^{p-1}-(k-F)^{p-1}\biggl\}(\Lambda N_r y+\Lambda^{-1}NR')\frac{F'}{F}   \nonumber \\
& + (n-2p-1)\biggl\{\left(k-F\right)^{p} +2pk^{p-1}F\biggl\}N\Lambda R^{n-2-2p}\Biggr],\label{I1}\\
I_2:=&\frac{(n-2)V_{n-2}^{(k)}}{2\kappa_n^2}\sum^{[n/2]}_{p=0}\int d^2{\bar y}p{\tilde \alpha}_{(p)} R^{n-2p-1}\biggl[-2k^{p-1}y {\dot \Lambda}-\{k^{p-1}-(k-F)^{p-1}\}\Lambda y\frac{\dot F}{F}\biggl].
\end{align}
In order to perform the variation of $I_M$, we have to deal with the terms containing $F'$ and ${\dot F}$.
Using the binomial expansion and integration by parts, we can rewrite the second line of Eq.~(\ref{I1}) as
\begin{align}
& \frac{(n-2)V_{n-2}^{(k)}}{2\kappa_n^2}\sum^{[n/2]}_{p=0}{\tilde \alpha}_{(p)} pR^{n-2p-1}\biggl\{k^{p-1}-(k-F)^{p-1}\biggl\}(\Lambda N_r y+\Lambda^{-1}NR')\frac{F'}{F}  \nonumber \\
=&-\frac{(n-2)V_{n-2}^{(k)}}{2\kappa_n^2}\sum^{[n/2]}_{p=2}{\tilde \alpha}_{(p)} pR^{n-2p-1}\sum_{w=0}^{p-2}\frac{(p-1)!(-1)^{p-1-w}}{w!(p-1-w)!}k^{w}F^{p-1-w}(\Lambda N_r y+\Lambda^{-1}NR')\frac{F'}{F}   \nonumber \\
=&-\frac{(n-2)V_{n-2}^{(k)}}{2\kappa_n^2}\sum^{[n/2]}_{p=2}{\tilde \alpha}_{(p)} p\sum_{w=0}^{p-2}\frac{(p-1)!(-1)^{p-1-w}k^{w}}{w!(p-1-w)!}\nonumber \\
&\times \biggl\{ \sum_{j=0}^{p-2-w}\frac{2(p-2-w)!(-1)^{p-2-w-j}}{j!(p-2-w-j)!}\frac{(N_r R^{n-2p-1}\Lambda^{1-2j}{R'}^{2j})'y^{2(p-w-j)-1}}{2(p-w-j)-1} \nonumber \\
&+\Lambda N_r yR^{n-2p-1}F^{p-2-w}(\Lambda^{-2}{R'}^2)' -(\Lambda^{-1}NR'R^{n-2p-1})'\frac{F^{p-1-w}}{p-1-w}\biggl\}.\label{app2-1}
\end{align}
This expression does not contain $y'$ and can be used to obtain $P_R$.
An important observation is that the only part of the action $I_M$  in (\ref{I_M}) that contributes to $P_\Lambda$ is $I_2$.  
After a tedious but straightforward calculation, using binomial expansion and integration by parts yet again, we can rewrite the last term in $I_2$ as
\begin{align}
&-\frac{(n-2)V_{n-2}^{(k)}}{2\kappa_n^2}\sum^{[n/2]}_{p=0}\int d^2{\bar y}p{\tilde \alpha}_{(p)} R^{n-2p-1}\biggl\{k^{p-1}-(k-F)^{p-1}\biggl\}\Lambda y\frac{\dot F}{F} \nonumber \\
=&-\frac{(n-2)V_{n-2}^{(k)}}{\kappa_n^2}\sum^{[n/2]}_{p=2}\int d^2{\bar y}p{\tilde \alpha}_{(p)} \sum_{w=1}^{p-1}\frac{(p-1)!(-1)^wk^{p-1-w}}{w!(p-1-w)!}R^{n-2p-1} F^{w-1} y\Lambda^{-2}{\dot \Lambda}{R'}^2 \nonumber \\
&+\frac{(n-2)V_{n-2}^{(k)}}{\kappa_n^2}\sum^{[n/2]}_{p=2}\int d^2{\bar y}p{\tilde \alpha}_{(p)} \sum_{w=1}^{p-1}\frac{(p-1)!(-1)^wk^{p-1-w}}{w!(p-1-w)!}\sum_{j=0}^{w-1}\frac{(w-1)!(-1)^{w-1-j}}{j!(w-1-j)!} \nonumber \\
&\times \biggl[\frac{1}{2(w-j)+1}\partial_t(R^{n-2p-1}\Lambda^{1-2j}){R'}^{2j}y^{2(w-j)+1} \nonumber \\
&- \sum_{q=0}^{2(w-j)+1}\frac{(2w-2j)!(-1)^q}{q!(2w-2j+1-q)!}\frac{j{\dot R}^{2(w-j+1)}}{w-j+1}(R^{n-2p-1}\Lambda^{1-2j}N^{-2(w-j)-1}{R'}^{2j-1+q}N_r^q)' \nonumber \\
&-\sum_{q=0}^{2(w-j)-1}\frac{(2w-2j-1)!(-1)^q}{q!(2w-2j-1-q)!}\frac{{\dot R}^{2w-2j-q}}{2w-2j-q}(R^{n-2p-1}\Lambda^{-1-2j}N^{-2(w-j)+1}{R'}^{2j+1+q}N_r^q)'\biggl] \nonumber \\
&+\partial_t(\cdots)+\partial_x(\cdots).\label{app2-2}
\end{align}
Using Eqs.~(\ref{app2-1}) and (\ref{app2-2}), we obtain the Lagrangian density (\ref{Lag-1}).

\subsection{Liouville form (\ref{Liouville}) in Lovelock gravity}
\label{app:Liouville}
In this appendix, we verify the Liouville form (\ref{Liouville}) in Lovelock gravity.
Note that the explicit form of $P_R$ is not used in the derivation.

Using Eqs.~(\ref{qlm-L}) and (\ref{eq:PM}), we write $P_M\delta M$ as
\begin{align}
P_M\delta M=&-\frac{(n-2){\cal A}_{n-2}}{2\kappa_n^2}\sum_{p=0}^{[n/2]}{\tilde \alpha}_{(p)}\frac{y\Lambda}{F}R^{n-2-2p}\biggl[(n-1-2p)(1-F)^p\delta R-pR(1-F)^{p-1}\delta F\biggl] \nonumber \\
=&-\frac{(n-2){\cal A}_{n-2}}{2\kappa_n^2}\sum_{p=0}^{[n/2]}{\tilde \alpha}_{(p)}(n-1-2p)\frac{y\Lambda}{F}R^{n-2-2p}(1-F)^p\delta R \nonumber \\
&+\frac{(n-2){\cal A}_{n-2}}{2\kappa_n^2}\sum_{p=1}^{[n/2]}{\tilde \alpha}_{(p)}R^{n-3}\frac{y\Lambda}{F}\delta F \nonumber \\
&+\frac{(n-2){\cal A}_{n-2}}{2\kappa_n^2}\sum_{p=2}^{[n/2]}{\tilde \alpha}_{(p)}pR^{n-1-2p}\sum_{w=1}^{p-1}\frac{(p-1)!(-1)^w}{w!(p-1-w)!}F^{w-1}y\Lambda\delta F.
\end{align} 
Using Eq.~(\ref{useful1}) for the second term and Eq.~(\ref{useful0}) for the last term together with the binomial expansion, we obtain
\begin{align}
P_M\delta M=&-\frac{(n-2){\cal A}_{n-2}}{2\kappa_n^2}\sum_{p=0}^{[n/2]}{\tilde \alpha}_{(p)}(n-1-2p)\frac{y\Lambda}{F}R^{n-2-2p}(1-F)^p\delta R \nonumber \\
& +\frac{(n-2){\cal A}_{n-2}}{2\kappa_n^2}\sum_{p=1}^{[n/2]}{\tilde \alpha}_{(p)}pR^{n-1-2p}\biggl\{2\delta(y\Lambda)-2y\delta \Lambda-R'\delta \ln\biggl|\frac{R'+y\Lambda}{R'-y\Lambda}\biggl|\biggl\} \nonumber \\
&+\frac{(n-2){\cal A}_{n-2}}{2\kappa_n^2}\sum_{p=2}^{[n/2]}{\tilde \alpha}_{(p)}\biggl[pR^{n-1-2p}\sum_{w=1}^{p-1}\frac{(p-1)!(-1)^w}{w!(p-1-w)!} \nonumber \\
&\times \sum_{j=0}^{w-1}\frac{(w-1)!(-1)^{w-1-j}}{j!(w-1-j)!}y^{2(w-1-j)}\biggl(\frac{{R'}^2}{\Lambda^2}\biggl)^{j}\biggl\{y\Lambda \biggl(\frac{\delta({R'}^2)}{\Lambda^2}-2y\delta y\biggl)-2y(y^2+F)\delta \Lambda\biggl\}\biggl]. \label{PMdeltaM1}
\end{align} 
An important fact is that $P_M\delta M$ has the form of $P_M\delta M=P_\Lambda\delta \Lambda+(\cdots)\delta R+\delta \eta+\zeta'$, where $\delta \eta$ and $\zeta'$ directly appear in the Liouville form (\ref{Liouville}), because $S$ is defined by $S:=R$.
Therefore all the terms with $\delta \Lambda$ in Eq. (A.16) are 
contained in the expression for $P_\Lambda$.
Using the binomial expansion and integration by parts, we can calculate the other terms in Eq.~(\ref{PMdeltaM1}) as follows
\begin{align}
&-\frac{(n-2){\cal A}_{n-2}}{2\kappa_n^2}\sum_{p=0}^{[n/2]}{\tilde \alpha}_{(p)}(n-1-2p)\frac{y\Lambda}{F}R^{n-2-2p}(1-F)^p\delta R \nonumber \\
& +\frac{(n-2){\cal A}_{n-2}}{2\kappa_n^2}\sum_{p=1}^{[n/2]}{\tilde \alpha}_{(p)}pR^{n-1-2p}\biggl\{2\delta(y\Lambda)-R'\delta \ln\biggl|\frac{R'+y\Lambda}{R'-y\Lambda}\biggl|\biggl\} \nonumber \\
&+\frac{(n-2){\cal A}_{n-2}}{2\kappa_n^2}\sum_{p=2}^{[n/2]}{\tilde \alpha}_{(p)}\biggl[pR^{n-1-2p}\sum_{w=1}^{p-1}\frac{(p-1)!(-1)^w}{w!(p-1-w)!} \nonumber \\
&\times \sum_{j=0}^{w-1}\frac{(w-1)!(-1)^{w-1-j}}{j!(w-1-j)!}y^{2(w-1-j)}\biggl(\frac{{R'}^2}{\Lambda^2}\biggl)^{j}y\Lambda \biggl(\frac{\delta({R'}^2)}{\Lambda^2}-2y\delta y\biggl)\biggl] \nonumber  \\
=&-\frac{(n-2){\cal A}_{n-2}}{2\kappa_n^2}\sum_{p=0}^{[n/2]}{\tilde \alpha}_{(p)}(n-1-2p)\frac{y\Lambda}{F}R^{n-2-2p}(1-F)^p\delta R \nonumber \\
& +\frac{(n-2){\cal A}_{n-2}}{2\kappa_n^2}\sum_{p=1}^{[n/2]}{\tilde \alpha}_{(p)}\delta \biggl[pR^{n-1-2p}\biggl\{2y\Lambda-R' \ln\biggl|\frac{R'+y\Lambda}{R'-y\Lambda}\biggl|\biggl\}\biggl] \nonumber \\
& -\frac{(n-2){\cal A}_{n-2}}{2\kappa_n^2}\sum_{p=1}^{[n/2]}{\tilde \alpha}_{(p)}p\biggl[2(n-1-2p)R^{n-2-2p}y\Lambda\delta R \nonumber \\
&-\biggl((n-1-2p)R^{n-2-2p}R'\delta R+R^{n-1-2p}\delta (R')\biggl)\ln\biggl|\frac{R'+y\Lambda}{R'-y\Lambda}\biggl|\biggl] \nonumber \\
&+\frac{(n-2){\cal A}_{n-2}}{\kappa_n^2}\sum_{p=2}^{[n/2]}{\tilde \alpha}_{(p)}pR^{n-1-2p}\sum_{w=1}^{p-1}\frac{(p-1)!(-1)^w}{w!(p-1-w)!}F^{w-1}\frac{yR'}{\Lambda}\delta(R') \nonumber\\
&-\frac{(n-2){\cal A}_{n-2}}{\kappa_n^2}\sum_{p=2}^{[n/2]}{\tilde \alpha}_{(p)}\delta \biggl[pR^{n-1-2p}\Lambda \sum_{w=1}^{p-1}\frac{(p-1)!(-1)^w}{w!(p-1-w)!} \nonumber \\
&\times \sum_{j=0}^{w-1}\frac{(w-1)!(-1)^{w-1-j}}{j!(w-1-j)!}\biggl(\frac{{R'}^2}{\Lambda^2}\biggl)^{j}\frac{y^{2(w-j)+1}}{2(w-j)+1}\biggl] \nonumber \\
&+\frac{(n-2){\cal A}_{n-2}}{\kappa_n^2}\sum_{p=2}^{[n/2]}{\tilde \alpha}_{(p)}\frac{py^{2(w-j)+1}}{2(w-j)+1}\delta\biggl[R^{n-1-2p}\Lambda \sum_{w=1}^{p-1}\frac{(p-1)!(-1)^w}{w!(p-1-w)!} \nonumber \\
&\times \sum_{j=0}^{w-1}\frac{(w-1)!(-1)^{w-1-j}}{j!(w-1-j)!}\biggl(\frac{{R'}^2}{\Lambda^2}\biggl)^{j}\biggl].
\end{align} 
Because there will not appear any more total variation terms, we can read off the total variation term $\eta$ as Eq.~(\ref{deta}).

In order to derive $\zeta$, we write down the quantity $\Pi:= P_M\delta M-P_\Lambda\delta\Lambda-\delta \eta$:
\begin{align}
\Pi=&-\frac{(n-2){\cal A}_{n-2}}{2\kappa_n^2}\sum_{p=0}^{[n/2]}{\tilde \alpha}_{(p)}(n-1-2p)\frac{y\Lambda}{F}R^{n-2-2p}(1-F)^p\delta R \nonumber \\
& -\frac{(n-2){\cal A}_{n-2}}{2\kappa_n^2}\sum_{p=1}^{[n/2]}{\tilde \alpha}_{(p)}p(n-1-2p)R^{n-2-2p}\biggl\{2y\Lambda -R'\ln\biggl|\frac{R'+y\Lambda}{R'-y\Lambda}\biggl|\biggl\}\delta R  \nonumber \\
& +\frac{(n-2){\cal A}_{n-2}}{2\kappa_n^2}\sum_{p=1}^{[n/2]}{\tilde \alpha}_{(p)}\biggl[pR^{n-1-2p}\ln\biggl|\frac{R'+y\Lambda}{R'-y\Lambda}\biggl|\delta R\biggl]' \nonumber \\
& -\frac{(n-2){\cal A}_{n-2}}{2\kappa_n^2}\sum_{p=1}^{[n/2]}{\tilde \alpha}_{(p)}\biggl[pR^{n-1-2p}\ln\biggl|\frac{R'+y\Lambda}{R'-y\Lambda}\biggl|\biggl]'\delta R \nonumber \\
&+\frac{(n-2){\cal A}_{n-2}}{2\kappa_n^2}\sum_{p=2}^{[n/2]}{\tilde \alpha}_{(p)}\biggl[2pR^{n-1-2p}\sum_{w=1}^{p-1}\frac{(p-1)!(-1)^w}{w!(p-1-w)!}F^{w-1}\frac{yR'}{\Lambda}\delta R \biggl]' \nonumber\\
&-\frac{(n-2){\cal A}_{n-2}}{2\kappa_n^2}\sum_{p=2}^{[n/2]}{\tilde \alpha}_{(p)}\biggl[2pR^{n-1-2p}\sum_{w=1}^{p-1}\frac{(p-1)!(-1)^w}{w!(p-1-w)!}F^{w-1}\frac{yR'}{\Lambda}\biggl]'\delta R  \nonumber\\
&+\frac{(n-2){\cal A}_{n-2}}{\kappa_n^2}\sum_{p=2}^{[n/2]}{\tilde \alpha}_{(p)}p(n-1-2p)R^{n-2-2p}\Lambda \sum_{w=1}^{p-1}\frac{(p-1)!(-1)^w}{w!(p-1-w)!} \nonumber \\
&\times \sum_{j=0}^{w-1}\frac{(w-1)!(-1)^{w-1-j}}{j!(w-1-j)!}\biggl(\frac{{R'}^2}{\Lambda^2}\biggl)^{j}\frac{y^{2(w-j)+1}}{2(w-j)+1}\delta R \nonumber \\
&+\frac{(n-2){\cal A}_{n-2}}{\kappa_n^2}\sum_{p=2}^{[n/2]}{\tilde \alpha}_{(p)}pR^{n-1-2p} \sum_{w=1}^{p-1}\frac{(p-1)!(-1)^w}{w!(p-1-w)!} \nonumber \\
&\times \sum_{j=0}^{w-1}\frac{(w-1)!(-1)^{w-1-j}}{j!(w-1-j)!}\frac{2j{R'}^{2j-1}\Lambda^{1-2j}}{2(w-j)+1}y^{2(w-j)+1}\delta (R').
\end{align} 
The last term generates a total derivative term by integration by parts.
Now we see all the total derivative terms and can read off the total variation term $\zeta$ to be Eq.~(\ref{zeta'}).

In order to prove the Liouville form (\ref{Liouville}), we write down the quantity $\Xi\delta R:= P_M\delta M-P_\Lambda\delta\Lambda-\delta \eta-\zeta'$ as
\begin{align}
\Xi\delta R=&-\frac{(n-2){\cal A}_{n-2}}{2\kappa_n^2}\sum_{p=0}^{[n/2]}{\tilde \alpha}_{(p)}(n-1-2p)\frac{y\Lambda}{F}R^{n-2-2p}(1-F)^p\delta R \nonumber \\
& -\frac{(n-2){\cal A}_{n-2}}{2\kappa_n^2}\sum_{p=1}^{[n/2]}{\tilde \alpha}_{(p)}pR^{n-2-2p}\biggl[2(n-1-2p)y\Lambda+R\biggl(\ln\biggl|\frac{R'+y\Lambda}{R'-y\Lambda}\biggl|\biggl)'\biggl] \delta R\nonumber \\
&+\frac{(n-2){\cal A}_{n-2}}{\kappa_n^2}\sum_{p=2}^{[n/2]}{\tilde \alpha}_{(p)}p(n-1-2p)R^{n-2-2p}\Lambda \sum_{w=1}^{p-1}\frac{(p-1)!(-1)^w}{w!(p-1-w)!} \nonumber \\
&\times \sum_{j=0}^{w-1}\frac{(w-1)!(-1)^{w-1-j}}{j!(w-1-j)!}\biggl(\frac{{R'}^2}{\Lambda^2}\biggl)^{j}\frac{y^{2(w-j)+1}}{2(w-j)+1}\delta R \nonumber \\
&-\frac{(n-2){\cal A}_{n-2}}{\kappa_n^2}\sum_{p=2}^{[n/2]}{\tilde \alpha}_{(p)}\biggl[pR^{n-1-2p}\sum_{w=1}^{p-1}\frac{(p-1)!(-1)^w}{w!(p-1-w)!} \nonumber \\
&\times \biggl\{F^{w-1}\frac{yR'}{\Lambda} + \sum_{j=0}^{w-1}\frac{(w-1)!(-1)^{w-1-j}}{j!(w-1-j)!}\frac{2j{R'}^{2j-1}\Lambda^{1-2j}}{2(w-j)+1}y^{2(w-j)+1}\biggl\}\biggl]'\delta R.
\end{align}

We now  show  that   
\begin{align}
\Xi=\frac{1}{R'}(\Lambda P_\Lambda'+P_MM'),
\end{align}
which is sufficient to verify the Liouville form (\ref{Liouville}).
From the expression (\ref{PLambda3}), we obtain
\begin{align}
\Lambda P_\Lambda'=&-\frac{(n-2){\cal A}_{n-2}}{\kappa_n^2}\sum_{p=1}^{[n/2]}{\tilde \alpha}_{(p)}p\Lambda \biggl\{(n-1-2p)R^{n-2-2p}yR'+R^{n-1-2p}y'\biggl\} \nonumber \\
&+\frac{(n-2){\cal A}_{n-2}}{2\kappa_n^2}\sum_{p=2}^{[n/2]}{\tilde \alpha}_{(p)}p(n-1-2p)R^{n-2-2p}R'\Lambda \nonumber \\
&\times  \biggl\{-2y(y^2+F)\sum_{w=1}^{p-1}\frac{(p-1)!(-1)^w}{w!(p-1-w)!}F^{w-1} \nonumber \\
&+\sum_{w=1}^{p-1}\frac{(p-1)!(-1)^w}{w!(p-1-w)!}\sum_{j=0}^{w-1}\frac{(w-1)!(-1)^{w-1-j}}{j!(w-1-j)!}\frac{2(1-2j)}{2(w-j)+1}y^{2(w-j)+1}(y^2+F)^{j}\biggl\} \nonumber \\
&+\frac{(n-2){\cal A}_{n-2}}{2\kappa_n^2}\sum_{p=2}^{[n/2]}{\tilde \alpha}_{(p)}pR^{n-1-2p}\Lambda \biggl\{-2y(y^2+F)\sum_{w=1}^{p-1}\frac{(p-1)!(-1)^w}{w!(p-1-w)!}F^{w-1} \nonumber \\
&+\sum_{w=1}^{p-1}\frac{(p-1)!(-1)^w}{w!(p-1-w)!}\sum_{j=0}^{w-1}\frac{(w-1)!(-1)^{w-1-j}}{j!(w-1-j)!}\frac{2(1-2j)}{2(w-j)+1}y^{2(w-j)+1}(y^2+F)^{j}\biggl\}'.
\end{align}  
Using 
\begin{align}
R'\biggl\{\ln\biggl|\frac{R'+y\Lambda}{R'-y\Lambda}\biggl|\biggl\}'=&2(y\Lambda)'-2y\Lambda'-y\Lambda \frac{F'}{F}
\end{align}
for the logarithmic term, we finally obtain
\begin{align}
\Xi\delta R=&-\frac{(n-2){\cal A}_{n-2}}{2\kappa_n^2}\sum_{p=0}^{[n/2]}{\tilde \alpha}_{(p)}(n-1-2p)\frac{y\Lambda}{F}R^{n-2-2p}(1-F)^p\delta R \nonumber \\
& -\frac{(n-2){\cal A}_{n-2}}{2\kappa_n^2}\sum_{p=1}^{[n/2]}{\tilde \alpha}_{(p)}\biggl[2p(n-1-2p)R^{n-2-2p}y\Lambda+pR^{n-1-2p}\frac{1}{R'}\biggl(2y'\Lambda-y\Lambda \frac{F'}{F}\biggl)\biggl] \delta R\nonumber \\
&+\frac{(n-2){\cal A}_{n-2}}{\kappa_n^2}\sum_{p=2}^{[n/2]}{\tilde \alpha}_{(p)}p(n-1-2p)R^{n-2-2p}\Lambda \sum_{w=1}^{p-1}\frac{(p-1)!(-1)^w}{w!(p-1-w)!} \nonumber \\
&\times \sum_{j=0}^{w-1}\frac{(w-1)!(-1)^{w-1-j}}{j!(w-1-j)!}\biggl(\frac{{R'}^2}{\Lambda^2}\biggl)^{j}\frac{y^{2(w-j)+1}}{2(w-j)+1}\delta R \nonumber \\
&-\frac{(n-2){\cal A}_{n-2}}{\kappa_n^2}\sum_{p=2}^{[n/2]}{\tilde \alpha}_{(p)}\biggl[pR^{n-1-2p}\sum_{w=1}^{p-1}\frac{(p-1)!(-1)^w}{w!(p-1-w)!} \nonumber \\
&\times \biggl\{F^{w-1}\frac{yR'}{\Lambda} + \sum_{j=0}^{w-1}\frac{(w-1)!(-1)^{w-1-j}}{j!(w-1-j)!}\frac{2j{R'}^{2j-1}\Lambda^{1-2j}}{2(w-j)+1}y^{2(w-j)+1}\biggl\}\biggl]'\delta R.
\end{align}
On the other hand, using Eqs.~(\ref{qlm-L}) and (\ref{eq:PM}), we obtain
\begin{align}
P_MM' =&-\frac{(n-2){\cal A}_{n-2}}{2\kappa_n^2}\sum_{p=0}^{[n/2]}{\tilde \alpha}_{(p)}(n-1-2p)R^{n-2-2p}(1-F)^pR'\frac{y\Lambda}{F} \nonumber \\
&+\frac{(n-2){\cal A}_{n-2}}{2\kappa_n^2}\sum_{p=1}^{[n/2]}{\tilde \alpha}_{(p)}pR^{n-1-2p}(1-F)^{p-1}F'\frac{y\Lambda}{F}.
\end{align}  
Some useful cancellations allow us to derive
\begin{align}
\Xi-\frac{\Lambda P_\Lambda+P_MM'}{R'}=& -\frac{(n-2){\cal A}_{n-2}}{2\kappa_n^2}\sum_{p=2}^{[n/2]}{\tilde \alpha}_{(p)}pR^{n-1-2p}\biggl[\frac{F'}{R'}\frac{y\Lambda}{F}\biggl\{(1-F)^{p-1}-1\biggl\} \nonumber \\
&+\sum_{w=1}^{p-1}\frac{(p-1)!(-1)^w}{w!(p-1-w)!} \biggl\{2yF^{w-1}(y^2+F)\biggl(\frac{\Lambda}{R'}\biggl)'+2F^{w-1}y^2y'\frac{\Lambda}{R'}\biggl\}  \nonumber \\
&+\sum_{w=1}^{p-1}\frac{(p-1)!(-1)^w}{w!(p-1-w)!} \sum_{j=0}^{w-1}\frac{(w-1)!(-1)^{w-1-j}}{j!(w-1-j)!}\frac{2jy^{2(w-j)+1}(y^2+F)^{j-1}}{2(w-j)+1}\nonumber \\
&\times \biggl\{2(y^2+F)\biggl(\frac{\Lambda}{R'}\biggl)' +(2yy'+F')\frac{\Lambda}{R'} \biggl\}\biggl].
\end{align}
Expanding the first term, we finally obtain
\begin{align}
&\Xi-\frac{\Lambda P_\Lambda+P_MM'}{R'} \nonumber \\
=& -\frac{(n-2){\cal A}_{n-2}}{2\kappa_n^2}\biggl\{2(y^2+F)\biggl(\frac{\Lambda}{R'}\biggl)' +(2yy'+F')\frac{\Lambda}{R'} \biggl\}\sum_{p=2}^{[n/2]}{\tilde \alpha}_{(p)}pR^{n-1-2p}\nonumber \\
&\times \sum_{w=1}^{p-1}\frac{(p-1)!(-1)^w}{w!(p-1-w)!}\biggl[F^{w-1}y+ \sum_{j=0}^{w-1}\frac{(w-1)!(-1)^{w-1-j}}{j!(w-1-j)!}\frac{2jy^{2(w-j)+1}(y^2+F)^{j-1}}{2(w-j)+1}\biggl].
\end{align}
By direct calculations, we can show
\begin{align}
2(y^2+F)\biggl(\frac{\Lambda}{R'}\biggl)' +(2yy'+F')\frac{\Lambda}{R'}  =0
\end{align}  
and complete the proof.

\subsection{Equation~(\ref{eq:GRfinal}) in Lovelock gravity}
\label{appendix3}
In this appendix, we show Eq.~(\ref{eq:GRfinal}) in Lovelock gravity.
{ While we consider the spherically symmetric case $(k=1$) in the maintext, we derive the equations for general $k$ in this appendix.}

Let us start from the action in the form of (\ref{I_M}):
\begin{align}
I_M=& \frac{(n-2)V_{n-2}^{(k)}}{2\kappa_n^2}\sum^{[n/2]}_{p=0}\int d^2{\bar y}{\tilde \alpha}_{(p)}\Biggl[2pk^{p-1}R^{n-2p-1}y(N_r\Lambda)' \nonumber \\
&-\frac{2pk^{p-1}N}{n-2p}\biggl((R^{n-2p})''\Lambda^{-1}+(R^{n-2p})'(\Lambda^{-1})'\biggl)  \nonumber \\
&+ pR^{n-2p-1}\biggl\{k^{p-1}-(k-F)^{p-1}\biggl\}\biggl\{(\Lambda N_r y+\Lambda^{-1}NR')\frac{F'}{F}-\Lambda y\frac{\dot F}{F}\biggl\}   \nonumber \\
& +(n-2p-1)\biggl\{\left(k-F\right)^{p} +2pk^{p-1}F\biggl\}N\Lambda R^{n-2-2p}-2pk^{p-1}R^{n-2p-1}y {\dot \Lambda}\biggl].
\end{align}
Using  
\begin{align}
(\Lambda^{-1})'=\frac{\Lambda(F'+2yy')}{2{R'}^2}-\frac{R''}{R'\Lambda}
\end{align}
for the second line and integration by parts for the first line, we obtain
\begin{align}
I_M=& \frac{(n-2)V_{n-2}^{(k)}}{2\kappa_n^2}\sum^{[n/2]}_{p=0}\int d^2{\bar y}{\tilde \alpha}_{(p)}\Biggl[-2pk^{p-1}\Lambda\biggl(N_r+\frac{Ny}{R'}\biggl)(R^{n-2p-1}y)' \nonumber \\
&+ pR^{n-2p-1}\biggl\{k^{p-1}-(k-F)^{p-1}\biggl\}\biggl\{\biggl(N_r+\frac{Ny}{R'}\biggl)\frac{\Lambda y}{F}F'-\Lambda y\frac{\dot F}{F}\biggl\}  \nonumber \\
& - pR^{n-2p-1}\frac{N\Lambda}{R'}(k-F)^{p-1}F' +(n-2p-1)\left(k-F\right)^{p}N\Lambda R^{n-2-2p} \nonumber \\
&-2pk^{p-1}R^{n-2p-1}y {\dot \Lambda}\biggl]+\mbox{(t.d.)},
\end{align}
where we also used ${R'}^2\Lambda^{-1}=F\Lambda+y^2\Lambda$.
Using 
\begin{align}
\frac{N\Lambda}{R'}M' =&\frac{(n-2)V_{n-2}^{(k)}}{2\kappa_n^2}\sum_{p=0}^{[n/2]}{\tilde \alpha}_{(p)}\frac{N\Lambda}{R'}R^{n-1-2p}\biggl[-p(k-F)^{p-1}F'+(n-1-2p)(k-F)^p\frac{R'}{R}\biggl],\\
P_M{\dot M} =&\frac{(n-2)V_{n-2}^{(k)}}{2\kappa_n^2}\sum_{p=0}^{[n/2]}{\tilde \alpha}_{(p)}\frac{y\Lambda}{F}R^{n-1-2p}\biggl[p(k-F)^{p-1}{\dot F}-(n-1-2p)(k-F)^p\frac{\dot R}{R}\biggl]
\end{align}  
which can be obtained from Eqs.~(\ref{qlm-L}) and (\ref{eq:PM}), we obtain
\begin{align}
&I_M-\int d^2{\bar y}\biggl(P_M{\dot M}+\frac{N\Lambda}{R'}M'\biggl) \nonumber \\
=& \frac{(n-2)V_{n-2}^{(k)}}{2\kappa_n^2}\sum^{[n/2]}_{p=0}\int d^2{\bar y}{\tilde \alpha}_{(p)}\Biggl[-2pk^{p-1}\Lambda\biggl(N_r+\frac{Ny}{R'}\biggl)(R^{n-2p-1}y)' \nonumber \\
&+ pR^{n-2p-1}\biggl\{k^{p-1}-(k-F)^{p-1}\biggl\}\biggl(N_r+\frac{Ny}{R'}\biggl)\frac{\Lambda y}{F}F'-2pk^{p-1}R^{n-2p-1}y {\dot \Lambda}  \nonumber \\
&-pk^{p-1}R^{n-2p-1}\Lambda y\frac{\dot F}{F} +\frac{y\Lambda}{F}(k-F)^p\partial_t(R^{n-1-2p})\biggl]+\mbox{(t.d.)}.
\end{align}
Using Eq.~(\ref{useful1}) for $y\Lambda{\dot F}/F$ together with integration by parts and ${\dot R}/R'=Ny/R'+N_r$, we rewrite the above expression as
\begin{align}
&I_M-\int d^2{\bar y}\biggl(P_M{\dot M}+\frac{N\Lambda}{R'}M'\biggl) \nonumber \\
=& \frac{(n-2)V_{n-2}^{(k)}}{2\kappa_n^2}\sum^{[n/2]}_{p=0}\int d^2{\bar y}{\tilde \alpha}_{(p)}\Biggl[-2pk^{p-1}\Lambda \frac{{\dot R}}{R'}(R^{n-2p-1}y)' \nonumber \\
&+ pR^{n-2p-1}\biggl\{k^{p-1}-(k-F)^{p-1}\biggl\}\frac{{\dot R}}{R'}\frac{\Lambda y}{F}F' +\frac{y\Lambda}{F}(k-F)^p\partial_t(R^{n-1-2p}) \nonumber \\
&+pk^{p-1}\biggl\{2y\Lambda\partial_t(R^{n-2p-1})+\frac{1}{n-2p}\partial_t(R^{n-2p})\partial_x\biggl(\ln\biggl|\frac{R'+y\Lambda}{R'-y\Lambda}\biggl|\biggl)\biggl\}\biggl]+\mbox{(t.d.)}.
\end{align}
Replacing the logarithmic term by 
\begin{align}
\partial_x\biggl(\ln\biggl|\frac{R'+y\Lambda}{R'-y\Lambda}\biggl|\biggl)=&\frac{-2R''y\Lambda+2(y\Lambda)'R'}{F\Lambda^2} \nonumber \\
=&\frac{y\Lambda}{FR'}\biggl(-F'+\frac{2y'}{y}F\biggl),
\end{align}
we finally obtain
\begin{align}
I_M-\int d^2{\bar y}\biggl(P_M{\dot M}+\frac{N\Lambda}{R'}M'\biggl) =& \frac{(n-2)V_{n-2}^{(k)}}{2\kappa_n^2}\sum^{[n/2]}_{p=0}\int d^2{\bar y}{\tilde \alpha}_{(p)}\frac{{\dot R}\Lambda yR^{n-2-2p}(k-F)^{p-1}}{R'F} \nonumber \\
&\times \Biggl[(n-2p-1)(k-F)R'- pRF'\biggl]+\mbox{(t.d.)}\nonumber \\
=& \int d^2{\bar y}\frac{{\dot R}\Lambda y}{R'F}M'+\mbox{(t.d.)}
\end{align}
and completes the derivation.

\section{Boundary condition at spacelike infinity}
\label{app:boundary}
In this appendix, we confirm that under the transformation from $\{\Lambda,P_\Lambda;R, P_R\}$ to $\{M, P_M;S, P_S\}$  the boundary term and total variation in the (\ref{Liouville}) are finite with suitable asymptotic fall-off rates of the ADM variables consistent with asymptotic flatness. 

We consider the following behavior of the ADM variables near spacelike infinity:
\begin{align}
&N\simeq N_\infty(t)+\mathcal{O}(x^{-\epsilon_1}),\\
&N_r\simeq N_r^{\infty}(t) x^{-(n-3)/2-\epsilon_2},\\
&\Lambda \simeq 1+\Lambda_1(t) x^{-(n-3)-\epsilon_3},\\
&R\simeq x+R_1(t) x^{-(n-4)-\epsilon_4},
\end{align}
where $\epsilon_1$ is positive and $\epsilon_2$--$\epsilon_4$ are non-negative numbers, and require the following three;\\
 (I) the canonical transformation from $\{\Lambda,P_\Lambda;R, P_R\}$ to $\{M, P_M;S, P_S\}$ is well-defined, \\
 (II) the Hamiltonian is finite in terms both of $\{\Lambda,P_\Lambda;R, P_R\}$ and $\{M, P_M;S, P_S\}$, and \\
 (III) the Misner-Sharp mass is non-zero finite at spacelike infinity $M\simeq M^\infty(t)$.
 
We will see below that these requirements are fulfilled for $\epsilon_1>0$, $\epsilon_2>\max[0,-(n-5)/2]$, $\epsilon_3= 0$, and $\epsilon_4>\max[0,-(n-5)]$, which are adopted as the boundary condition in the present paper. 

First we see the following integrated Liouville form (\ref{Liouville}):
\begin{align}
\int_{-\infty}^{\infty} dx(P_\Lambda\delta \Lambda+P_R\delta R)-\int_{-\infty}^{\infty} dx(P_M\delta M+P_S\delta S)=\delta\int_{-\infty}^{\infty}\eta dx +[~\zeta~]^{x=\infty}_{x=-\infty},\label{intLiouv}
\end{align} 
where $P_\Lambda$, $P_R$, $S$, $P_S$, $M$, $P_M$, $\eta$, and $\zeta$ are defined by Eqs.~(\ref{PLambdaGR2}), (\ref{PR}), (\ref{eq:R}), (\ref{eq:PM}), (\ref{MS}), (\ref{eq:Pr1}), (\ref{deta-gr}), and (\ref{zeta'-gr}), respectively.
For the well-definedness of the canonical transformation, two conditions must hold at spacelike infinity; (i) $\zeta$ vanishes, and (ii) the integrands in the left-hand side and $\eta$ converge to zero faster than $\mathcal{O}(x^{-1})$.
The second requirement ensures the finiteness of the integrals.

Let us see the asymptotic behavior of the momentum conjugates.
Near spacelike infinity, $P_\Lambda$ and $P_R$ behave as
\begin{align}
P_\Lambda\simeq& -\frac{(n-2)\ma A_{n-2}}{\kappa_n^2}N_\infty^{-1}\biggl({\dot R_1} x^{1-\epsilon_4}-N_r^{\infty} x^{(n-3)/2-\epsilon_2}\biggl), \\
P_R\simeq & -\frac{(n-2)\ma A_{n-2} }{\kappa_n^2} N_\infty^{-1}\biggl[N_r^{\infty}(t) \biggl(-\frac{n-3}{2}+\epsilon_2\biggl)x^{(n-5)/2-\epsilon_2}+{\dot \Lambda}_1(t) x^{-\epsilon_3}+(n-3){\dot R_1} x^{-\epsilon_4}\biggl],
\end{align}
with which we obtain
\begin{align}
P_\Lambda\delta\Lambda\simeq& \mathcal{O}(x^{-(n-4)-\epsilon_3-\epsilon_4})+\mathcal{O}(x^{-(n-3)/2-\epsilon_2-\epsilon_3}), \\
P_R\delta R\simeq & \mathcal{O}(x^{-(n-3)/2-\epsilon_2-\epsilon_4})+\mathcal{O}(x^{-(n-4)-\epsilon_3-\epsilon_4})+\mathcal{O}(x^{-(n-4)-2\epsilon_4}).
\end{align}
Hence, $\epsilon_4>0$ and $\epsilon_2+\epsilon_3>0$ are required for $n=5$.
For $n=4$, the requirement is $\epsilon_4>1/2$, $\epsilon_3+\epsilon_4>1$, $\epsilon_2+\epsilon_4>1/2$, and $\epsilon_2+\epsilon_3>1/2$.

For the next check, we see the behavior of $F$, defined by Eq.~(\ref{defF}).
Using 
\begin{align}
y \simeq&N_\infty^{-1}({\dot R}_1x^{-(n-4)-\epsilon_4}-N_r^{\infty} x^{-(n-3)/2-\epsilon_2}),\label{eval-y}
\end{align}
where $y$ is defined by Eq.~(\ref{defy}), we obtain
\begin{align}
F\simeq &N_\infty^{-2}{\dot R}_1x^{-(n-4)}(-{\dot R}_1x^{-(n-4)-2\epsilon_4}+2N_r^{\infty}x^{-(n-3)/2-\epsilon_2-\epsilon_4}) \nonumber \\
&+1-2\Lambda_1 x^{-(n-3)-\epsilon_3}-N_\infty^{-2}{N_r^{\infty}}^2x^{-(n-3)-2\epsilon_2}-2(n-4+\epsilon_4)R_1x^{-(n-3)-\epsilon_4},\label{asympF}
\end{align} 
which converges to $1$.
For the condition (III) required, $F$ must behave near infinity as 
\begin{align}
F\simeq 1-F_1(t)x^{-(n-3)}
\end{align} 
and then, for $M\simeq M^\infty(t)$,  $F_1$ is identified as
\begin{align}
F_1(t)\equiv \frac{2\kappa_n^2M^\infty(t)}{(n-2)\ma A_{n-2}}.
\end{align} 
Since we have already required $(n-4)+2\epsilon_4>1$ and $(n-3)/2+\epsilon_2+\epsilon_4>1$ in the previous argument, the first line in (\ref{asympF}) converges to zero faster than $\mathcal{O}(x^{-(n-3)})$.
Therefore the condition (III) requires $\epsilon_2\epsilon_3\epsilon_4=0$, where $F_1(t)$ is determined depending on the cases.

Using Eq.~(\ref{eval-y}) and $M\simeq M^\infty(t)$, we obtain the asymptotic behavior of $P_M\delta M$ as
\begin{align}
P_M\delta M\simeq -N_\infty^{-1}\delta M^\infty({\dot R}_1x^{-(n-4)-\epsilon_4}-N_r^{\infty} x^{-(n-3)/2-\epsilon_2}).
\label{eq:PM2}
\end{align}
This provides additional requirements; $\epsilon_2>0$ for $n=5$ and $\epsilon_4>1$ and $\epsilon_2>1/2$ for $n=4$.
{ Note however that if one uses the momentum variable $\tilde{P}_M$ appropriate for asymptotically PG slices, then there is an extra term on the right hand side of (\ref{eq:PM2}) that cancels the second term. This in turn allows the choice $\epsilon_2=0$ for all spacetime dimensions as required by the PG slicing.}

The conditions for $\epsilon_2$ and $\epsilon_4$ obtained up to here are summarized as $\epsilon_2>\max[0,-(n-5)/2]$ and $\epsilon_4>\max[0,-(n-5)]$.
We will see below that the requirements (I) and (II) are fulfilled under these conditions.

Using the following asymptotic expansion of $P_S$;  
\begin{align}
P_S\simeq -\frac{(n-2)\ma A_{n-2} }{\kappa_n^2} N_\infty^{-1}\biggl({\dot \Lambda}_1 x^{-\epsilon_3}+(n-4+\epsilon_4){\dot R_1} x^{-\epsilon_4}\biggl)+\mathcal{O}(x^{-(n-1)/2-\epsilon_2-\epsilon_M}), \label{asympPS} 
\end{align}
where $\epsilon_M$ is some positive number defined by the next leading-order of $M$ as $M\simeq M^\infty+\mathcal{O}(x^{-\epsilon_M})$, we obtain
\begin{align}
P_S\delta S \simeq & -\delta R_1\frac{(n-2)\ma A_{n-2} }{\kappa_n^2} N_\infty^{-1}\biggl({\dot \Lambda}_1 x^{-(n-4)-\epsilon_3-\epsilon_4}+(n-4+\epsilon_4){\dot R_1} x^{-(n-4)-2\epsilon_4}\biggl) \nonumber \\
&+\mathcal{O}(x^{-3(n-3)/2-\epsilon_2-\epsilon_4-\epsilon_M}).
\label{eq:eta2}
\end{align}
Under the present conditions this converges to zero faster than $\mathcal{O}(x^{-1})$.
{ Here too one should keep in mind that with the choice $\epsilon_2=0$ needed for PG slicings, there is an extra total variation arising from the transformation from $P_{M}$ to $\tilde{P}_M$. This extra term cancels the second term on the right hand side of (\ref{eq:eta2}) rendering $\eta$ finite for $\epsilon_2=0$ in all dimensions.}

Next let us evaluate $\zeta$ and $\eta$.
We write the logarithmic term in the expressions (\ref{deta-gr}) and (\ref{zeta'-gr}) as
\begin{align}
\ln\biggl(\frac{R'+y\Lambda}{R'-y\Lambda}\biggl)=&\ln\biggl(\frac{1-W}{1+W}\biggl),
\end{align} 
where
\begin{align} 
W:=&\frac{\kappa_n^2\Lambda P_\Lambda}{(n-2)\ma A_{n-2}R^{n-3}R' }.
\end{align} 
Using the fact that $W$ converges to zero as
\begin{align}
W\simeq -N_\infty^{-1}\biggl({\dot R_1} x^{-(n-4)-\epsilon_4}-N_r^{\infty} x^{-(n-3)/2-\epsilon_2}\biggl),
\end{align}
we can evaluate the logarithmic term as
\begin{align}
\ln\biggl(\frac{R'+y\Lambda}{R'-y\Lambda}\biggl)\simeq &-2W-\frac23 W^3.
\end{align} 
Using this, we can show that $\zeta$ converges to zero under the present conditions as
\begin{align}
\zeta\simeq &\mathcal{O}(x^{-(n-5)-2\epsilon_4})+\mathcal{O}(x^{-(n-5)/2-\epsilon_2-\epsilon_4}).
\end{align} 
On the other hand, $\eta$ is evaluated as
\begin{align}
\eta=&\frac{(n-2)\ma A_{n-2}R^{n-3}R' }{2\kappa_n^2}\biggl[2W+\ln\biggl(\frac{1-W}{1+W}\biggl)\biggl] \nonumber \\
\simeq &-\frac{(n-2)\ma A_{n-2}x^{n-3}}{3\kappa_n^2}W^3 \nonumber \\
\simeq &\mathcal{O}(x^{-2(n-4)+1-3\epsilon_4})~~\mbox{or}~~\mathcal{O}(x^{-(n-3)/2-3\epsilon_2})
\end{align} 
and converges to zero faster than $\mathcal{O}(x^{-1})$ under the present conditions.

Lastly, let us check the well-definedness of the Hamiltonian $\int^\infty_{-\infty}dx(NH+N_rH_r)$, where $H_r$ and $H$ are defined by (\ref{Hr}) and (\ref{H}), respectively.
It requires that the fall-off rates of $NH$ and $N_rH_r$ are faster than $\mathcal{O}(x^{-1})$.
From the following asymptotic expressions;
\begin{align}
\Lambda P_\Lambda'\simeq& -\frac{(n-2)\ma A_{n-2}}{\kappa_n^2}N_\infty^{-1}\biggl(1+\Lambda_1 x^{-(n-3)-\epsilon_3}\biggl) \nonumber \\
&\times \biggl[(1-\epsilon_4){\dot R_1} x^{-\epsilon_4}-\biggl(\frac{n-3}{2}-\epsilon_2\biggl)N_r^{\infty} x^{(n-5)/2-\epsilon_2}\biggl], \\
R'P_R\simeq & -\frac{(n-2)\ma A_{n-2} }{\kappa_n^2} N_\infty^{-1}\biggl(1-(n-4+\epsilon_4)x^{-(n-3)-\epsilon_4}\biggl) \nonumber \\
&\times \biggl[N_r^{\infty}(t) \biggl(-\frac{n-3}{2}+\epsilon_2\biggl)x^{(n-5)/2-\epsilon_2}+{\dot \Lambda}_1(t) x^{-\epsilon_3}+(n-3){\dot R_1} x^{-\epsilon_4}\biggl], 
\end{align}
we see that the dangerous terms of the order $x^{(n-5)/2-\epsilon_2}$ in $H_r=-\Lambda P_\Lambda'+R'P_R$ are canceled out and obtain
\begin{align}
N_rH_r\simeq&  \mathcal{O}(x^{-(n-3)/2-\epsilon_2-\epsilon_3})+\mathcal{O}(x^{-(n-3)/2-\epsilon_2-\epsilon_4}).
\end{align}
This is faster than $\mathcal{O}(x^{-1})$ under the present conditions.
Next let us see the behavior of $NH$.
Using the followings asymptotic expansions;
\begin{align}
\Lambda P_\Lambda \simeq& -\frac{(n-2)\ma A_{n-2}}{\kappa_n^2}N_\infty^{-1}\biggl(1+\Lambda_1x^{-(n-3)-\epsilon_3}\biggl)\biggl({\dot R_1} x^{1-\epsilon_4}-N_r^{\infty} x^{(n-3)/2-\epsilon_2}\biggl), \\
RP_R\simeq & -\frac{(n-2)\ma A_{n-2} }{\kappa_n^2} N_\infty^{-1}\biggl(x+R_1x^{-(n-4)-\epsilon_4}\biggl) \nonumber \\
&\times \biggl[N_r^{\infty}(t) \biggl(-\frac{n-3}{2}+\epsilon_2\biggl)x^{(n-5)/2-\epsilon_2}+{\dot \Lambda}_1(t) x^{-\epsilon_3}+(n-3){\dot R_1} x^{-\epsilon_4}\biggl], 
\end{align}
we obtain 
\begin{align}
\frac{P_{\Lambda}}{R^{n-2}}\biggl(RP_{R}-\frac{n-3}{2}\Lambda P_{\Lambda}\biggl) \simeq &\mathcal{O}(x^{-(n-4)-\epsilon_3-\epsilon_4})+ \mathcal{O}(x^{-(n-3)/2-\epsilon_2-\epsilon_4})+\mathcal{O}(x^{-(n-3)/2-\epsilon_2-\epsilon_3}) \nonumber \\
&+ 
\mathcal{O}(x^{-(n-4)-2\epsilon_4}).
\end{align} 
This shows that the first line of the following super-Hamiltonian;
\begin{align}
H=&-\frac{\kappa_n^2P_{\Lambda}}{(n-2)\ma A_{n-2}R^{n-2}}\biggl(RP_{R}-\frac{n-3}{2}\Lambda P_{\Lambda}\biggl) \nonumber \\
&-\frac{(n-2)\ma A_{n-2}}{\kappa_n^2}\biggl\{-R^{n-3}(R' \Lambda^{-1})'+\frac{n-3}{2} \Lambda R^{n-4}(1-\Lambda^{-2}{R'}^2)\biggl\}
\end{align} 
converges to zero faster than $\mathcal{O}(x^{-1})$ under the present conditions.
On the other hand, the second line is evaluated as
\begin{align}
-R^{n-3}(R' \Lambda^{-1})'+\frac{n-3}{2} \Lambda R^{n-4}(1-\Lambda^{-2}{R'}^2) \simeq \mathcal{O}(x^{-1-\epsilon_3})+\mathcal{O}(x^{-1-\epsilon_4}),
\end{align} 
which also converges faster than $\mathcal{O}(x^{-1})$.
As a result, $NH$ converge to zero faster than $\mathcal{O}(x^{-1})$ and hence the Hamiltonian is well-defined.

We also check the well-definedness of the Hamiltonian with a new set of variables: $\int^\infty_{-\infty}dx(N^M M'+N^S P_S)$, where $N^M$ and $N^S$ are defined by (\ref{NM}) and (\ref{NS}), respectively.
Using Eq.~(\ref{eval-y}), we obtain $N^M\simeq N_\infty$.
Combined this with $M'\simeq \mathcal{O}(x^{-1-\epsilon_M})$, it is shown that $N^MM'$ converges to zero faster than $\mathcal{O}(x^{-1})$ under the present conditions.
On the other hand, using $N^S\simeq \mathcal{O}(x^{-(n-4)-\epsilon_4})$ and Eq.~(\ref{asympPS}), we obtain
\begin{align}
N^SP_S\simeq& \mathcal{O}(x^{-(n-4)-\epsilon_3-\epsilon_4})+\mathcal{O}(x^{-(n-4)-2\epsilon_4})+\mathcal{O}(x^{-3(n-3)/2-\epsilon_2-\epsilon_4-\epsilon_M}).
\end{align}
This also converges to zero faster than $\mathcal{O}(x^{-1})$ under the present conditions.


\begin{thebibliography}{99}
\bibitem{BHuniqueness}
P.T.~Chrusciel, J.L.~Costa, and M.~Heusler,
Living Rev. Rel. {\bf 15}, 7 (2012).
\bibitem{er2008}
R.~Emparan and H.S.~Reall,
Living Rev. Rel. {\bf 11}, 6 (2008).
\bibitem{myersperry1986}
R. C. Myers, M. J. Perry, {Annals of Phys.} {\bf 172}, 304 (1986).
\bibitem{er2002}
R.~Emparan and H.S.~Reall,
Phys.\ Rev.\ Lett.\  {\bf 88}, 101101 (2002).
\bibitem{ida2000}
D.~Ida,
Phys. Rev. Lett. {\bf 85}, 3758 (2000).
\bibitem{Lovelock} 
D. Lovelock, J. Math. Phys. {\bf 12}, 498 (1971).
\bibitem{lanczos}
C.~Lanczos,
Annals Math. {\bf 39}, 842 (1938).
\bibitem{string} 
For a recent review, see S. Wadia,``String Theory: A Framework for Quantum Gravity and Various Applications'', TWAS Jubilee Publication, [arXiv:hep-th/0809.1036v2].
\bibitem{string lovelock} 
   D. J.~Gross and E.~Witten, 
   Nucl. Phys. {\bf B277}, 1 (1986);\\
   D. J.~Gross and J. H.~Sloan, 
   Nucl. Phys. {\bf B291}, 41 (1987);\\
   R. R.~Metsaev and A. A.~Tseytlin, 
Phys. Lett. B {\bf 191}, 354 (1987);\\
  B.~Zwiebach,
   Phys. Lett. B {\bf 156}, 315 (1985);\\
   R. R.~Metsaev and A. A.~Tseytlin, 
   Nucl. Phys. {\bf B293}, 385 (1987).

\bibitem{lovelockreview} 
  C.~Garraffo and G.~Giribet,
  Mod.\ Phys.\ Lett.\  A {\bf 23}, 1801 (2008);\\
C.~Charmousis,
  Lect.\ Notes Phys.\  {\bf 769}, 299 (2009).
\bibitem{kuchar94} 
K. Kucha\v{r}, 
Phys. Rev. {\bf D50}, 3961 (1994).
\bibitem{Louko1996}
J.~Louko and J.~M\"akel\"a, 
Phys. Rev. D {\bf 54}, 4982 (1996).
\bibitem{Bekenstein}
J.D.~Bekenstein, 
Lett. al Nuovo Cimento {\bf 11}, 467 (1974); \\
V.~Mukhanov, JETP Letters {\bf 44}, 66 (1986);\\
J.D.~Bekenstein and V.F.~Mukhanov, 
Phys. Lett. {\bf B360}, 7 (1995).
\bibitem{area spectrum}  
A.~Barvinsky, S.~Das, and G.~Kunstatter,  
Phys. Lett. {\bf B517}, 415 (2001);\\
H.A.~Kastrup, 
Phys. Lett. {\bf B385}, 75 (1996).
\bibitem{whitt1988} 
B.~Whitt, 
Phys. Rev. D {\bf 38}, 3000 (1988). 
\bibitem{ms1964} 
C. W.~Misner and D. H.~Sharp, 
Phys. Rev. {\bf 136}, B571 (1964).
\bibitem{hayward1996}
S. A. Hayward, 
Phys. Rev. D. {\bf 53}, 1938 (1996).
\bibitem{zegers2005} 
R.~Zegers, 
J. Math. Phys. {\bf 46}, 072502 (2005). 
\bibitem{mwr2011}
H.~Maeda, S.~Willison, and S.~Ray, 
Class. Quant. Grav. {\bf 28}, 165005 (2011).
\bibitem{HM08}
H. Maeda and M. Nozawa, 
Phys. Rev. D {\bf 77} 064031 (2008).
\bibitem{GTMletter} 
G.~Kunstatter, T.~Taves, and H.~Maeda,
Class. Quant. Grav. {\bf 29}, 092001 (2012).
\bibitem{TZ} 
C. Teitelboim and J. Zanelli, 
Class. Quant. Grav. {\bf 4}, L125 (1987).
\bibitem{df2012} 
S.~Deser and J.~Franklin,
Class. Quant. Grav. {\bf 29},  072001 (2012).
\bibitem{st2008} 
T.~Torii and H.~Shinkai,
Phys. Rev. D {\bf 78}, 084037 (2008).
\bibitem{JL97}
J. Louko, J. Simon and S. Winters-Hilt, 
Phys. Rev. D {\bf 55} 3526 (1997).
\bibitem{TLKM} 
T. Taves, C. D. Leonard, G. Kunstatter and R.B. Mann, 
Class. Quant. Grav. {\bf 29}, 015012 (2012).
\bibitem{wald}
R.M.~Wald, {\it General Relativity}, (University of Chicago Press, 1984).
\bibitem{olea2007}
G.~Kofinas and R.~Olea,
JHEP {\bf 0711}, 069 (2007).
\bibitem{tw2011}
Yu~Tian and X.-N.~Wu,
JHEP {\bf 1101}, 150 (2011).
\bibitem{bdw} 
D.G.~Boulware and S.~Deser,
Phys.\ Rev.\ Lett.\ \textbf{55}, 2656 (1985).
\bibitem{bdw2} 
D.~G. Boulware and S.~Deser,
Phys. Lett. {\bf B175}, 409 (1986);\\
J.~T.~Wheeler,
Nucl.\ Phys.\ \textbf{B268}, 737 (1986);\\
D.~Lorenz-Petzold, 
Mod. Phys. Lett. {\bf A3}, 827 (1988);\\
R.-G.~Cai,
Phys. Rev. D {\bf 65}, 084014 (2002);\\
R.-G.~Cai and Qi~Guo,
Phys. Rev. D {\bf 69}, 104025 (2004).
\bibitem{DCBH}
  M.~Ba\~nados, C.~Teitelboim and J.~Zanelli,
  Phys.\ Rev.\  D {\bf 49}, 975 (1994);\\
R.-G.~Cai and K.-S.~Soh,
Phys. Rev. D {\bf 59}, 044013 (1999).
\bibitem{DCBH2}
J.~Crisostomo, R.~Troncoso, and J.~Zanelli,
Phys. Rev. D {\bf 62}, 084013 (2000);\\
R.~Aros, R.~Troncoso, and J.~Zanelli,
Phys. Rev. D {\bf 63}, 084015 (2001).
\bibitem{cd2002} 
C.~Charmousis and J-F.~Dufaux, 
Class.\ Quant.\ Grav.\ {\bf 19}, 4671 (2002).
\bibitem{Garraffo:2007fi}
  C.~Garraffo, G.~Giribet, E.~Gravanis and S.~Willison,
  J.\ Math.\ Phys.\  {\bf 49}, 042502 (2008).
\bibitem{Gravanis:2010zs}
E.~Gravanis, 
Phys. Rev. D {\bf 82}, 104024 (2010).
\bibitem{Palais1979} R.S. Palais, Comm.\ Math.\ Phys.\ {\bf 69}, 19 (1979).
\bibitem{Fels2002} M.E. Fels and C.G. Torre, Class.\ Quant.\ Grav.\ {\bf 19}, 641 (2002).
\bibitem{Deser2003} S. Deser and B. Tekin,
Class.\ Quant.\ Grav.\ {\bf 20}, 4877 (2003)
\bibitem{Louko2007} 
G. Kunstatter and J. Louko, 
Phys. Rev. D {\bf 75}, 024036 (2007).
\bibitem{Husain} 
V. Husain and O. Winkler, 
Phys. Rev. D {\bf 71}, 104001 (2005).
\bibitem{KMT12} 
G. Kunstatter, H. Maeda, and T. Taves, in preparation.





\end{thebibliography}
\end{document}